\documentclass[12pt]{article}
\usepackage{graphicx}

\def\hybrid{\topmargin 0pt      \oddsidemargin 0pt
        \headheight 0pt \headsep 0pt
       \voffset-1cm
        \textwidth 6.25in       
       \textheight 9.5in       
        \marginparwidth 0.0in
        \parskip 5pt plus 1pt   \jot = 1.5ex}
\catcode`\@=11
\def\marginnote#1{}

\newcount\hour
\newcount\minute
\newtoks\amorpm
\hour=\time\divide\hour by60
\minute=\time{\multiply\hour by60 \global\advance\minute by-\hour}
\edef\standardtime{{\ifnum\hour<12 \global\amorpm={am}%
        \else\global\amorpm={pm}\advance\hour by-12 \fi
        \ifnum\hour=0 \hour=12 \fi
        \number\hour:\ifnum\minute<10 0\fi\number\minute\the\amorpm}}
\edef\militarytime{\number\hour:\ifnum\minute<10 0\fi\number\minute}

\def\draftlabel#1{{\@bsphack\if@filesw {\let\thepage\relax
   \xdef\@gtempa{\write\@auxout{\string
      \newlabel{#1}{{\@currentlabel}{\thepage}}}}}\@gtempa
   \if@nobreak \ifvmode\nobreak\fi\fi\fi\@esphack}
        \gdef\@eqnlabel{#1}}
\def\@eqnlabel{}
\def\@vacuum{}
\def\draftmarginnote#1{\marginpar{\raggedright\scriptsize\tt#1}}

\def\draftlabel#1{{\@bsphack\if@filesw {\let\thepage\relax
   \xdef\@gtempa{\write\@auxout{\string
      \newlabel{#1}{{\@currentlabel}{\thepage}}}}}\@gtempa
   \if@nobreak \ifvmode\nobreak\fi\fi\fi\@esphack}
        \gdef\@eqnlabel{#1}}
\def\@eqnlabel{}
\def\@vacuum{}
\def\draftmarginnote#1{\marginpar{\raggedright\scriptsize\tt#1}}

\def\draft{\oddsidemargin -.5truein
        \def\@oddfoot{\sl preliminary draft \hfil
        \rm\thepage\hfil\sl\today\quad\militarytime}
        \let\@evenfoot\@oddfoot \overfullrule 3pt
        \let\label=\draftlabel
        \let\marginnote=\draftmarginnote
   \def\@eqnnum{(\theequation)\rlap{\kern\marginparsep\tt\@eqnlabel}%
\global\let\@eqnlabel\@vacuum}  }

\def\numberbysection{\@addtoreset{equation}{section}
        \def\theequation{\thesection.\arabic{equation}}}

\def\underline#1{\relax\ifmmode\@@underline#1\else
        $\@@underline{\hbox{#1}}$\relax\fi}

\def\titlepage{\@restonecolfalse\if@twocolumn\@restonecoltrue\onecolumn
     \else \newpage \fi \thispagestyle{empty}\c@page\z@
        \def\thefootnote{\fnsymbol{footnote}} }

\def\endtitlepage{\if@restonecol\twocolumn \else  \fi
        \def\thefootnote{\arabic{footnote}}
        \setcounter{footnote}{0}}  
\relax

\numberbysection
\hybrid

\newfont{\Bbb}{msbm10 scaled 1\@ptsize00}
\newfont{\Bbbb}{msbm7 scaled 1\@ptsize00}

\newcommand{\DDD}{\raise-1pt\hbox{$\mbox{\Bbbb D}$}}

\newcommand{\UUU}{\raise-1pt\hbox{$\mbox{\Bbbb U}$}}

\newcommand{\ZZ}{\mbox{\Bbb Z}}
\newcommand{\z}{\raise-1pt\hbox{$\mbox{\Bbbb Z}$}}

\newcommand{\sss}{\raise-1pt\hbox{$\mbox{\Bbbb S}$}}

\def\beq{\begin{equation}}
\def\eeq{\end{equation}}
\def\p{\partial}

\newtheorem{lemma-definition}{Lemma-Definition}[section]

\def\normord{ {\scriptstyle {{\bullet}\atop{\bullet}}} }
\def\lbr{\left <}
\def\rbr{\right >}

\begin{document}

\begin{titlepage}

\title{Multicomponent DKP hierarchy and its dispersionless limit}

\author{A. Savchenko\thanks{
Skolkovo Institute of Science and Technology, 143026, Moscow, Russia
and National Research University Higher School of Economics,
20 Myasnitskaya Ulitsa,
Moscow 101000, Russia,
e-mail: ksanoobshhh@gmail.com}
\and
A.~Zabrodin\thanks{
Skolkovo Institute of Science and Technology, 143026, Moscow, Russia and
National Research University Higher School of Economics,
20 Myasnitskaya Ulitsa,
Moscow 101000, Russia and
NRC ``Kurchatov institute'', Moscow, Russia;
e-mail: zabrodin@itep.ru}}

\date{July 2024}
\maketitle

\vspace{-7cm} \centerline{ \hfill ITEP-TH-23/24}\vspace{7cm}

\begin{abstract}

Using the free fermions technique and bosonization rules 
we introduce the multicomponent
DKP hierarchy as a generating bilinear integral equation for the 
tau-function. A number of bilinear equations of the Hirota-Miwa type
are obtained as its corollaries. We also consider the dispersionless 
version of the hierarchy as a set of nonlinear differential equations
for the dispersionless limit of logarithm of the tau-function 
(the $F$-function). We show that there is an elliptic curve built
in the structure of the hierarchy, with the elliptic modulus being 
a dynamical variable. This curve can be uniformized by 
elliptic functions, and in the elliptic parametrization many 
dispersionless equations of the Hirota-Miwa type become equivalent
to a single equation having a nice form.

\end{abstract}

\end{titlepage}

\vspace{5mm}

%

\tableofcontents

\vspace{5mm}

\section{Introduction}

Integrable hierarchies of nonlinear differential equations such as
Kadomtsev-Petviashvili (KP) and Toda lattice are well studied at 
the present time. In the approach developed by the Kyoto school
\cite{DJKM83,JM83} the universal dependent variable of the hierarchies,
the tau-function, is represented as a vacuum expectation value of certain 
operators constructed from free fermions. 
These operators are exponential functions of neutral quadratic forms
in the free fermions $\psi_n, \psi^*_n$:
\beq\label{int1}
g=\exp \Bigl (\sum_{n,m \in \z}A_{nm}\psi_n\psi^*_m\Bigr ).
\eeq
Such operators obey the basic bilinear identity
\beq\label{int2}
\sum_{j\in \z}
\psi_{j}g \otimes \psi_{j}^*g 
=\sum_{j\in \z}
g\psi_{j}\otimes g \psi_{j}^*
\eeq
which, after the bosonization procedure, gives rise to bilinear 
equations for the tau-function. 

A similar approach for more general operators, which are exponential
functions of general (not necessarily neutral) quadratic forms,
leads to the DKP (also known as Pfaff lattice or coupled KP equation) 
and Pfaff-Toda
hierarchies, see \cite{HO}--\cite{Takasaki09}.

In a more general case, the fermionic operators can carry an additional
index $\alpha$: $\psi_n^{(\alpha )}, \psi^{*(\alpha )}_n$
(multicomponent fermions), $\alpha =1, \ldots , N$. 
The corresponding theory is a source of matrix and 
multicomponent hierarchies; for neutral quadratic forms they are
multicomponent KP and Toda hierarchies \cite{DJKM81}--\cite{TT07}.
In this paper we consider exponential functions of 
general quadratic expressions in the multicomponent fermions of the form 
\beq\label{int3}
g=\exp \left ( \sum_{\alpha , \beta}
\sum_{j,k}\Bigl ( A_{jk}^{(\alpha \beta )}
\psi^{(\alpha )}_{j}\psi^{*(\beta )}_{k}+
B_{jk}^{(\alpha \beta )}
\psi^{(\alpha )}_{j}\psi^{(\beta )}_{k}+
C_{jk}^{(\alpha \beta )}
\psi^{*(\alpha )}_{j}\psi^{*(\beta )}_{k}\Bigr )
\right ).
\eeq
Such operators obey the following bilinear relation:
\beq\label{int4}
\sum_{\gamma =1}^N \sum_{j\in \z}\Bigl (
\psi_{j}^{(\gamma )}g \otimes
\psi_{j}^{*(\gamma )}g +
\psi_{j}^{*(\gamma )}g \otimes
\psi_{j}^{(\gamma )}g \Bigr )
=\sum_{\gamma =1}^N \sum_{j\in \z}\Bigl (
g\psi_{j}^{(\gamma )}\otimes g \psi_{j}^{*(\gamma )}+
g\psi_{j}^{*(\gamma )}\otimes g \psi_{j}^{(\gamma )}\Bigr ).
\eeq
After bosonization it gives rise to bilinear equations for the 
tau-function. It is natural to call this hierarchy the multicomponent
DKP hierarchy.

The aim of this paper is two-fold. First, we introduce the multicomponent
DKP hierarchy in terms of its tau-function and find the 
generating bilinear equation for it using the free fermions technique. 
Second, we take the 
dispersionless limit of this hierarchy and represent the dispersionless 
equations in elliptic parametrization in which they considerably
simplify and acquire a nice form. 

The general approach to dispersionless 
integrable hierarchies was developed in \cite{TT95} by Takasaki and Takebe.
In \cite{TT07}, they analyzed the dispersionless limit of the 
multicomponent KP hierarchy.
We obtain nonlinear equations for the dispersionless limit
of (logarithm of) tau-function (the $F$-function) of the multicomponent
DKP hierarchy
which follow from the bilinear
equations of the Hirota-Miwa type. 

It turns out that in the dispersionless
limit of the DKP hierarchy there is an elliptic curve built 
in its structure \cite{Takasaki07,Takasaki09}, 
with the elliptic modulus being a dynamical variable. Uniformizing
this curve via elliptic functions, one can represent the equations
in a nice elliptic form (\ref{main}) (for the one-component case see \cite{AZ14}). 

The structure of the paper is as follows. In section 2 
we start from introducing
the multicomponent fermions, define the tau-function as expectation 
value of certain fermionic operators and obtain
bilinear equations for it. Next we consider their dispersionless limit,
uncover the elliptic curve and introduce the uniformization of this curve
in terms of elliptic functions (section 3). 
In the elliptic parametrization, many equations for the $F$-function
reduce to a nice single equation. Section 4 contains concluding 
remarks. There are also four appendices devoted to some technical 
details.

\section{Multicomponent DKP hierarchy}

\subsection{The multicomponent fermions}

Following \cite{DJKM81,TT07}, 
we introduce the creation-annihilation multicomponent free 
fermionic operators $\psi_{j}^{(\alpha )}$, 
$\psi_{j}^{*(\alpha )}$ ($j\in \ZZ$, $\alpha =1, \ldots , N$).
They obey the anti-commutation relations
$$
[\psi_{j}^{(\alpha )}, \psi_{k}^{*(\beta )}]_+=\delta_{\alpha \beta}\delta_{jk},
\qquad
[\psi_{j}^{(\alpha )}, \psi_{k}^{(\beta )}]_+=
[\psi_{j}^{*(\alpha )}, \psi_{k}^{*(\beta )}]_+=0.
$$
We also introduce 
free fermionic fields
$$
\psi^{(\alpha )}(z)=\sum_{j\in \z}\psi^{(\alpha )}_j z^j,
\qquad
\psi^{*(\alpha )}(z)=\sum_{j\in \z}\psi^{*(\alpha )}_j z^{-j}.
$$

The Fock and dual Fock spaces are generated by the vacuum states 
$\left | {\bf 0}\rbr$, $\lbr {\bf 0} \right |$ that satisfy the conditions
$$
\psi_{j}^{(\alpha )}\left | {\bf 0}\rbr =0 \quad (j<0), \qquad
\psi_{j}^{*(\alpha )}\left | {\bf 0}\rbr =0 \quad (j\geq 0),
$$
$$
\lbr {\bf 0}\right | \psi_{j}^{(\alpha )} =0 \quad (j\geq 0), \qquad
\lbr {\bf 0}\right | \psi_{j}^{*(\alpha )} =0 \quad (j< 0),
$$
so $\psi_{j}^{(\alpha )}$ with $j<0$ and $\psi_{j}^{*(\alpha )}$ with
$j\geq 0$ are annihilation operators while 
$\psi_{j}^{(\alpha )}$ with $j\geq 0$ and
$\psi_{j}^{*(\alpha )}$ with
$j<0$ are creation operators. 
Let ${\bf n}=(n_1, n_2, \ldots , n_N)$ be a set
of integer numbers. We define the states 
$\left | {\bf n}\rbr$, $\lbr {\bf n} \right |$
as
$$
\left | {\bf n}\rbr =\Psi_{n_N}^{*(N)}\ldots \Psi_{n_2}^{*(2)}
\Psi_{n_1}^{*(1)}\left | {\bf 0}\rbr , \qquad
\lbr {\bf n} \right |=\lbr {\bf 0} \right |\Psi_{n_1}^{(1)}\Psi_{n_2}^{(2)}\ldots
\Psi_{n_N}^{(N)},
$$
where
$$
\Psi_{n}^{*(\alpha )}=\left \{ \begin{array}{l}
\psi^{(\alpha )}_{n-1}\ldots \psi^{(\alpha )}_{0} \quad \,\,\, (n >0)
\\
\psi^{*(\alpha )}_{n}\ldots \psi^{*(\alpha )}_{-1} \quad (n <0),
\end{array}
\right.
$$
$$
\Psi_{n}^{(\alpha )}=\left \{ \begin{array}{l}
\psi^{*(\alpha )}_{0}\ldots \psi^{*(\alpha )}_{n-1} \quad  \, (n >0)
\\
\psi^{(\alpha )}_{-1}\ldots \psi^{(\alpha )}_{n} \quad \,\,\,\,\, (n <0).
\end{array}
\right.
$$

Let us introduce the operators
$$
J_{k}^{(\alpha )}=\sum_{j\in \z}\normord \psi^{(\alpha )}_{j} \psi^{*(\alpha )}_{j+k}
\normord ,
$$
where the normal ordering is defined by moving the annihilation operators 
to the right and creation operators to the left with the minus sign emerging each time
when two fermionic operators are permuted. 
Let
$$
{\bf t}=\{{\bf t}_1, {\bf t}_2, \ldots , {\bf t}_N\}, \qquad
{\bf t}_{\alpha}=\{t_{\alpha , 1}, t_{\alpha , 2}, t_{\alpha , 3}, \ldots \, \},
\qquad \alpha = 1, \ldots , N
$$
be $N$ infinite sets of the independent continuous time
variables. 
We introduce the operator
$$
J({\bf t})=\sum_{\alpha =1}^N \sum_{k\geq 1} t_{\alpha , k}J_k^{(\alpha )}.
$$
Note the commutation relations
\beq\label{comm}
e^{J({\bf t})}\psi^{(\gamma )}(z)=e^{\xi ({\bf t}_{\gamma}, z)}
\psi^{(\gamma )}(z)e^{J({\bf t})}, \qquad
e^{J({\bf t})}\psi^{*(\gamma )}(z)=e^{-\xi ({\bf t}_{\gamma}, z)}
\psi^{*(\gamma )}(z)e^{J({\bf t})},
\eeq
where
\beq\label{f5}
\xi ({\bf t}_{\gamma}, z)=\sum_{k\geq 1}t_{\gamma , k}z^k.
\eeq

\subsection{The bilinear identity}

The Clifford group element is exponent 
of a quadratic form in the fermionic operators 
$\psi_{j}^{(\alpha )}$, 
$\psi_{j}^{*(\alpha )}$. Its general form is
$$
g=\exp \left ( \sum_{\alpha , \beta}
\sum_{j,k}\Bigl ( A_{jk}^{(\alpha \beta )}
\psi^{(\alpha )}_{j}\psi^{*(\beta )}_{k}+
B_{jk}^{(\alpha \beta )}
\psi^{(\alpha )}_{j}\psi^{(\beta )}_{k}+
C_{jk}^{(\alpha \beta )}
\psi^{*(\alpha )}_{j}\psi^{*(\beta )}_{k}\Bigr )
\right )
$$
with some infinite matrices $A_{jk}^{(\alpha \beta )}$, 
$B_{jk}^{(\alpha \beta )}$, $C_{jk}^{(\alpha \beta )}$.
An important property of the Clifford group elements is the following 
operator bilinear identity:
\beq\label{f2}
\sum_{\gamma =1}^N \sum_{j\in \z}\Bigl (
\psi_{j}^{(\gamma )}g \otimes
\psi_{j}^{*(\gamma )}g +
\psi_{j}^{*(\gamma )}g \otimes
\psi_{j}^{(\gamma )}g \Bigr )
=\sum_{\gamma =1}^N \sum_{j\in \z}\Bigl (
g\psi_{j}^{(\gamma )}\otimes g \psi_{j}^{*(\gamma )}+
g\psi_{j}^{*(\gamma )}\otimes g \psi_{j}^{(\gamma )}\Bigr ).
\eeq
The proof can be found in Appendix A. 
In terms of the free fermionic fields
the operator bilinear identity acquires the form
\beq\label{f3}
\begin{array}{c}
\displaystyle{
\sum_{\gamma =1}^N \oint \frac{dz}{z} \Bigl (
\psi^{(\gamma )}(z)g\otimes \psi^{*(\gamma )}(z)g
+\psi^{*(\gamma )}(z)g\otimes \psi^{(\gamma )}(z)g\Bigr )}
\\ \\
\displaystyle{
=
\sum_{\gamma =1}^N \oint \frac{dz}{z} \Bigl ( 
g\psi^{(\gamma )}(z)\otimes g\psi^{*(\gamma )}(z)
+g\psi^{*(\gamma )}(z)\otimes g\psi^{(\gamma )}(z)\Bigr )}.
\end{array}
\eeq
Here the contour integral is understood to be an integral along the big circle
$|z|=R$ with sufficiently large $R$, $\oint dz z^n =2\pi i \delta_{n, -1}$.

Note that it holds
$$
\psi_{j}^{(\gamma )}\left |{\bf 0}\rbr \otimes \psi_{j}^{*(\gamma )}
\left |{\bf 0}\rbr =
\psi_{j}^{*(\gamma )}\left |{\bf 0}\rbr \otimes \psi_{j}^{(\gamma )}
\left |{\bf 0}\rbr =
0
$$
for all $j\in \ZZ$ because the vacuum $\left |{\bf 0}\rbr$ is annihilated
by either $\psi_{j}^{(\gamma )}$ or
$\psi_{j}^{*(\gamma )}$. 
Therefore, applying
both sides of (\ref{f2}) to 
$\left |{\bf 0}\rbr \otimes \left |{\bf 0}\rbr$, we get
$$
\sum_{\gamma =1}^N \sum_{j\in \z}\Bigl (
\psi_{j}^{(\gamma )}g \left |{\bf 0}\rbr \otimes
\psi_{j}^{*(\gamma )}g \left |{\bf 0}\rbr 
+\psi_{j}^{*(\gamma )}g \left |{\bf 0}\rbr \otimes
\psi_{j}^{(\gamma )}g \left |{\bf 0}\rbr \Bigr )
=0
$$
or
\beq\label{f4}
\sum_{\gamma =1}^N \oint \frac{dz}{z} \Bigl (
\psi^{(\gamma )}(z)g\left |{\bf 0}
\rbr \otimes \psi^{*(\gamma )}(z)g
\left |{\bf 0}\rbr 
+\psi^{*(\gamma )}(z)g\left |{\bf 0}
\rbr \otimes \psi^{(\gamma )}(z)g
\left |{\bf 0}\rbr \Bigr )
=0.
\eeq
Now apply $\lbr 2{\bf n}+{\bf e}_{\alpha}\right |e^{J({\bf t})}\otimes 
\lbr 2{\bf n}'-{\bf e}_{\beta}\right |e^{J({\bf t}')}$ from the left
to get
$$
\begin{array}{c}
\displaystyle{
\sum_{\gamma =1}^N \oint \frac{dz}{z}  
\lbr 2{\bf n}\! +\! {\bf e}_{\alpha}\right |e^{J({\bf t})}
\psi^{(\gamma )}(z)g\left |{\bf 0}\rbr 
\lbr 2{\bf n}'\! -\! {\bf e}_{\beta}\right |e^{J({\bf t}')}
\psi^{*(\gamma )}(z)g\left |{\bf 0}\rbr }
\\ \\
\displaystyle{
+\sum_{\gamma =1}^N \oint \frac{dz}{z}  
\lbr 2{\bf n}\! +\! {\bf e}_{\alpha}\right |e^{J({\bf t})}
\psi^{*(\gamma )}(z)g\left |{\bf 0}\rbr 
\lbr 2{\bf n}'\! -\! {\bf e}_{\beta}\right |e^{J({\bf t}')}
\psi^{(\gamma )}(z)g\left |{\bf 0}\rbr }
=0
\end{array}
$$
or
\beq\label{f6}
\begin{array}{c}
\displaystyle{
\sum_{\gamma =1}^N \oint \frac{dz}{z} \, 
e^{\xi ({\bf t}_{\gamma}-{\bf t}_{\gamma}', z)} 
\lbr 2{\bf n}\! +\! {\bf e}_{\alpha} \right | 
\psi^{(\gamma )}(z)e^{J({\bf t})}
g\left |{\bf 0}\rbr 
\lbr 2{\bf n}'\! -\! {\bf e}_{\beta} \right | 
\psi^{*(\gamma )}(z) e^{J({\bf t}')}
g\left |{\bf 0}\rbr }
\\ \\
\displaystyle{
+\sum_{\gamma =1}^N \oint \frac{dz}{z} \, 
e^{\xi ({\bf t}_{\gamma}-{\bf t}_{\gamma}', z)} 
\lbr 2{\bf n}\! +\! {\bf e}_{\alpha} \right | 
\psi^{*(\gamma )}(z)e^{J({\bf t})}
g\left |{\bf 0}\rbr 
\lbr 2{\bf n}'\! -\! {\bf e}_{\beta} \right | 
\psi^{(\gamma )}(z) e^{J({\bf t}')}
g\left |{\bf 0}\rbr  =0,}
\end{array}
\eeq
where the
commutation relations (\ref{comm}) are used.

\subsection{The generating 
bilinear functional relation for the tau-func\-tion}

The tau-function of the multicomponent  
DKP hierarchy is a $N\! \times \! N$ 
matrix function $\tau_{\alpha \beta} ({\bf n}, {\bf t})$ 
defined as the expectation value
\beq\label{f1}
\tau_{\alpha \beta} ({\bf n}, {\bf t})=
\lbr 2{\bf n}+{\bf e}_{\alpha} -
{\bf e}_{\beta}\right | e^{J({\bf t})} g\left |{\bf 0}\rbr ,
\eeq
where ${\bf e}_{\alpha}$ is the $N$-component vector whose $\alpha$'s
component is 1 and all other components are equal to 0 and 
$g$ is an element of the Clifford group. 
Note that $\lbr {\bf n}\right | e^{J({\bf t})} g\left |{\bf 0}\rbr =0$ 
unless $\displaystyle{\sum_{\alpha}n_{\alpha}}$ is an even
number; this suggests the choice of the left vacuum state in (\ref{f1}).
Note also that $\tau_{\alpha \alpha}({\bf n}, {\bf t})=:
\tau ({\bf n}, {\bf t})$ does not depend on the index $\alpha$.

In order to transform (\ref{f6}) to a functional relation 
for the tau-function,
we employ the multicomponent bosonization rules \cite{KL93}
\beq\label{bos}
\begin{array}{l}
\lbr {\bf n}\! +\! {\bf e}_{\alpha} \right | 
\psi^{(\gamma )}(z)e^{J({\bf t})}=
\epsilon_{\alpha \gamma}\epsilon_{\gamma}({\bf n})
z^{n_{\gamma}+\delta_{\alpha \gamma} -1}
\lbr {\bf n}\! +\! {\bf e}_{\alpha}\! -\! {\bf e}_{\gamma}\right |
e^{J({\bf t}-[z^{-1}]_{\gamma})},
\\ \\
\lbr {\bf n}\! -\! {\bf e}_{\beta} \right | \psi^{*(\gamma )}(z)
e^{J({\bf t})}=
\epsilon_{\beta \gamma}\epsilon_{\gamma}({\bf n})
z^{-n_{\gamma}+\delta_{\beta \gamma}}
\lbr {\bf n}\! -\! {\bf e}_{\beta}\! +\! {\bf e}_{\gamma}\right |
e^{J({\bf t}+[z^{-1}]_{\gamma})},
\end{array}
\eeq
where
$$
\left ({\bf t}\pm [z^{-1}]_{\gamma}\right )_{\alpha k}=t_{\alpha , k}\pm
\delta_{\alpha \gamma} \frac{z^{-k}}{k}
$$
and the sign factors $\epsilon_{\alpha \beta}$, $\epsilon_{\gamma}({\bf n})$ are:
$\epsilon_{\alpha \beta}=1$ if $\alpha \leq \beta$, 
$\epsilon_{\alpha \beta}=-1$ if $\alpha > \beta$,
$\epsilon_{\gamma}({\bf n})=(-1)^{n_{\gamma +1}+\ldots + n_N}$.
Note that $\epsilon_{\gamma}(2{\bf n})=1$.

Using the bosonization rules, we can rewrite (\ref{f6}) as
\beq\label{f7}
\begin{array}{l}
\displaystyle{
\sum_{\gamma} \epsilon_{\alpha \gamma}\epsilon_{\beta \gamma}
\oint \! dz \, z^{2n_{\gamma}-2n_{\gamma}' +\delta_{\alpha \gamma}
+\delta_{\beta \gamma}-2} e^{\xi ({\bf t}_{\gamma}-{\bf t}'_{\gamma}, z)}
\tau_{\alpha \gamma}({\bf n}, {\bf t}-[z^{-1}]_{\gamma})
\tau_{\gamma \beta}({\bf n}', {\bf t}'+[z^{-1}]_{\gamma})}
\\ \\
\displaystyle{
+\sum_{\gamma} \epsilon_{\alpha \gamma}\epsilon_{\beta \gamma}
\oint \! dz \, z^{2n_{\gamma}'-2n_{\gamma} -\delta_{\alpha \gamma}
-\delta_{\beta \gamma}-2} e^{-\xi ({\bf t}_{\gamma}-{\bf t}'_{\gamma}, z)}}
\\ \\
\phantom{aaaaaaaaaaaaaaaaaaaaaaa}\displaystyle{
\times \,
\tau_{\beta \gamma}({\bf n}'-{\bf e}_{\beta}, {\bf t}'-[z^{-1}]_{\gamma})
\tau_{\gamma \alpha}({\bf n}+{\bf e}_{\alpha}, {\bf t}+[z^{-1}]_{\gamma})}
=0.
\end{array}
\eeq
This is the bilinear functional relation for the tau-function valid
for all $\alpha , \beta$, ${\bf n}, {\bf n}', {\bf t}, {\bf t}'$.
The integration contour around $\infty$ is such that all singularities 
coming from the power of $z$ and the exponential function
$e^{\xi ({\bf t}_{\gamma}-{\bf t}_{\gamma}', \, z)}$ are inside it and all singularities 
coming from the $\tau$-factors are outside it. 

\subsection{Corollaries of the bilinear functional relation
(Hirota-Miwa equations)}

Choosing 
${\bf n}-{\bf n}'$, ${\bf t}-{\bf t}'$ in (\ref{f7}) in some special
ways and calculating the integrals using the residue calculus, 
one can obtain important bilinear equations for the tau-function
$\tau ({\bf n}, {\bf t})$.

First of all we consider (\ref{f7}) at 
$\beta =\alpha$:
\beq\label{f9}
\begin{array}{l}
\displaystyle{
\sum_{\gamma} 
\oint \! dz \, z^{2n_{\gamma}-2n_{\gamma}' +2\delta_{\alpha \gamma}
-2} e^{\xi ({\bf t}_{\gamma}-{\bf t}'_{\gamma}, z)}
\tau_{\alpha \gamma}({\bf n}, {\bf t}-[z^{-1}]_{\gamma})
\tau_{\gamma \alpha}({\bf n}', {\bf t}'+[z^{-1}]_{\gamma})}
\\ \\
\displaystyle{
+\sum_{\gamma} 
\oint \! dz \, z^{2n_{\gamma}'-2n_{\gamma} -2\delta_{\alpha \gamma}
-2} e^{-\xi ({\bf t}_{\gamma}-{\bf t}'_{\gamma}, z)}}
\\ \\
\phantom{aaaaaaaaaaaaaaaaaaaaaaa}\displaystyle{
\times \,
\tau_{\alpha \gamma}({\bf n}'-{\bf e}_{\alpha}, {\bf t}'-[z^{-1}]_{\gamma})
\tau_{\gamma \alpha}({\bf n}+{\bf e}_{\alpha}, {\bf t}+[z^{-1}]_{\gamma})}
=0.
\end{array}
\eeq
Put ${\bf n}'={\bf n}+{\bf e}_{\alpha}$ 
and ${\bf t}-{\bf t}'=[a^{-1}]_{\alpha}+
[b^{-1}]_{\alpha}$, so that
$$
e^{\xi ({\bf t}_{\alpha}-{\bf t}'_{\alpha}, z)}=
\frac{a\, b}{(a-z)(b-z)}, \quad
e^{\xi ({\bf t}_{\gamma}-{\bf t}'_{\gamma}, z)}=1 \quad
\mbox{for $\gamma \neq \alpha$}.
$$
The residue calculus and some transformations yield the equation
\beq\label{f10}
\begin{array}{c}
\displaystyle{
a^2 \frac{\tau ({\bf n}+{\bf e}_{\alpha}, {\bf t}+[b^{-1}]_{\alpha})
\tau ({\bf n}, {\bf t}+[a^{-1}]_{\alpha})}{\tau ({\bf n}+{\bf e}_{\alpha}, {\bf t}+[a^{-1}]_{\alpha}+[b^{-1}]_{\alpha})\tau ({\bf n}, {\bf t})}-
b^2 \frac{\tau ({\bf n}+{\bf e}_{\alpha}, {\bf t}+[a^{-1}]_{\alpha})
\tau ({\bf n}, {\bf t}+[b^{-1}]_{\alpha})}{\tau ({\bf n}+{\bf e}_{\alpha}, {\bf t}+[a^{-1}]_{\alpha}+[b^{-1}]_{\alpha})\tau ({\bf n}, {\bf t})}}
\\ \\
\displaystyle{=\, a^2-b^2 -(a-b)\p_{t_{\alpha , 1}}\log
\frac{\tau ({\bf n}+{\bf e}_{\alpha}, {\bf t}+[a^{-1}]_{\alpha}+[b^{-1}]_{\alpha})}{\tau ({\bf n}, {\bf t})}}.
\end{array}
\eeq
Next, differentiate (\ref{f9}) with respect to $t_{\alpha , 1}$ and
put ${\bf n}'={\bf n}$,
${\bf t}-{\bf t}'=[a^{-1}]_{\alpha}+
[b^{-1}]_{\alpha}$ after that. The residue calculus yields:
\beq\label{f11}
\begin{array}{c}
\displaystyle{
\frac{\tau ({\bf n}, {\bf t})\tau ({\bf n}, {\bf t}+[a^{-1}]_{\alpha}+[b^{-1}]_{\alpha})}{\tau ({\bf n}, {\bf t}+[a^{-1}]_{\alpha})
\tau ({\bf n}, {\bf t}+[b^{-1}]_{\alpha})}-
\frac{1}{a^2b^2}\,
\frac{\tau ({\bf n}-{\bf e}_{\alpha}, {\bf t})
\tau ({\bf n}+{\bf e}_{\alpha}, 
{\bf t}+[a^{-1}]_{\alpha}+[b^{-1}]_{\alpha})}{\tau ({\bf n}, {\bf t}+[a^{-1}]_{\alpha})
\tau ({\bf n}, {\bf t}+[b^{-1}]_{\alpha})}}
\\ \\
\displaystyle{
=\, 1-\frac{1}{a-b}\, \p_{t_{\alpha , 1}}\log
\frac{\tau ({\bf n}, {\bf t}+[a^{-1}]_{\alpha})}{\tau ({\bf n}, {\bf t}+[b^{-1}]_{\alpha})}}.
\end{array}
\eeq
These are ``diagonal'' equations which involve variables 
corresponding to one component only.
For each component, they coincide with the Hirota-Miwa 
equations of the one-component
DKP hierarchy. 
At $\alpha \neq \beta$, the integral bilinear equation (\ref{f7})
generates equations of the Hirota-Miwa type which mix different 
components. 

Let us now choose in (\ref{f9})
${\bf n}'={\bf n}$,
${\bf t}-{\bf t}' =[a^{-1}]_{\alpha}+[b^{-1}]_{\beta}$, where $\beta \neq \alpha$.
The residue calculus yields:
\beq\label{f111}
\begin{array}{c}
a\tau ({\bf n}, {\bf t}+[b^{-1}]_{\beta})
\tau ({\bf n}, {\bf t}+[a^{-1}]_{\alpha})-
a\tau ({\bf n}, {\bf t})
\tau ({\bf n}, {\bf t}+[a^{-1}]_{\alpha}+[b^{-1}]_{\beta})
\\ \\
+b^{-1}\tau_{\alpha \beta}({\bf n}, {\bf t}+[a^{-1}]_{\alpha})
\tau_{\beta \alpha}(({\bf n}, {\bf t}+[b^{-1}]_{\beta})
\\ \\
-
b^{-1}\tau_{\alpha \beta}({\bf n}-{\bf e}_{\alpha}, {\bf t})
\tau_{\beta \alpha}({\bf n}+{\bf e}_{\alpha}, {\bf t}
+[a^{-1}]_{\alpha}+[b^{-1}]_{\beta})=0.
\end{array}
\eeq

Now we consider (\ref{f7}) for $\alpha \neq \beta$. We start with
equations which do not contain derivatives with respect to continuous
variables. Put ${\bf n}'={\bf n}$, ${\bf t}-{\bf t}'=[a^{-1}]_{\alpha}+
[b^{-1}]_{\alpha}$. Equation (\ref{f7}) acquires the form
$$
\epsilon_{\beta \alpha}\oint dz \frac{ab}{z(a-z)(b-z)}\,
\tau ({\bf n}, {\bf t}-[z^{-1}]_{\alpha})\tau_{\alpha \beta}
({\bf n}, {\bf t}-[a^{-1}]_{\alpha}-
[b^{-1}]_{\alpha}+[z^{-1}]_{\alpha})
$$
$$
+\, \epsilon_{\alpha \beta}\oint dz z^{-1}
\tau_{\alpha \beta}({\bf n}, {\bf t}-[z^{-1}]_{\beta})
\tau ({\bf n}, {\bf t}-[a^{-1}]_{\alpha}-
[b^{-1}]_{\alpha}+[z^{-1}]_{\beta})
$$
$$
+\, \epsilon_{\beta \alpha}\oint dz \frac{(a-z)(b-z)}{abz^3}\,
\tau_{\beta \alpha} ({\bf n}-{\bf e}_{\beta}, {\bf t}-[a^{-1}]_{\alpha}-
[b^{-1}]_{\alpha}-[z^{-1}]_{\alpha})
\tau ({\bf n}+{\bf e}_{\alpha}, {\bf t}+[z^{-1}]_{\alpha})=0,
$$
where we write only non-zero terms in the sum over $\gamma$. 
The residue calculus followed by shift of the times yields:
\beq\label{f12}
\begin{array}{l}
a\tau ({\bf n}, {\bf t}+[a^{-1}]_{\alpha})
\tau_{\alpha \beta}({\bf n}, {\bf t}+[b^{-1}]_{\alpha})-
b\tau ({\bf n}, {\bf t}+[b^{-1}]_{\alpha})
\tau_{\alpha \beta}({\bf n}, {\bf t}+[a^{-1}]_{\alpha})
\\ \\
\phantom{aaaaaa}=\, (a-b)\tau ({\bf n}, {\bf t})
\tau_{\alpha \beta}({\bf n}, {\bf t}+[a^{-1}]_{\alpha}+[b^{-1}]_{\alpha})
\\ \\
\phantom{aaaaaaaaaaaaaaaaa}
+(a^{-1}-b^{-1})\tau ({\bf n}+{\bf e}_{\alpha},
{\bf t}+[a^{-1}]_{\alpha}+[b^{-1}]_{\alpha})\tau_{\alpha \beta}
({\bf n}-{\bf e}_{\alpha}, {\bf t}),
\end{array}
\eeq
where we took take into account that
\beq\label{f13}
\tau_{\beta \alpha}({\bf n}, {\bf t})=
\tau_{\alpha \beta}({\bf n}+{\bf e}_{\beta}-{\bf e}_{\alpha}, {\bf t}).
\eeq
Next we put ${\bf n}'={\bf n}+{\bf e}_{\beta}$, 
${\bf t}-{\bf t}'=[a^{-1}]_{\alpha}+ [b^{-1}]_{\alpha}$. In a similar way,
the residue calculus yields the equation
\beq\label{f14}
\begin{array}{l}
a\tau ({\bf n}-{\bf e}_{\beta}, {\bf t}+[a^{-1}]_{\alpha})
\tau_{\alpha \beta}({\bf n}, {\bf t}+[b^{-1}]_{\alpha})-
b\tau ({\bf n}-{\bf e}_{\beta}, {\bf t}+[b^{-1}]_{\alpha})
\tau_{\alpha \beta}({\bf n}, {\bf t}+[a^{-1}]_{\alpha})
\\ \\
\phantom{aaaaaa}=\, (a-b)\tau ({\bf n}-{\bf e}_{\beta}, {\bf t})
\tau_{\alpha \beta}({\bf n}, {\bf t}+[a^{-1}]_{\alpha}+[b^{-1}]_{\alpha})
\\ \\
\phantom{aaaaaaaaaaaaaaaaa}
+(a^{-1}-b^{-1})\tau ({\bf n}+{\bf e}_{\alpha}-{\bf e}_{\beta},
{\bf t}+[a^{-1}]_{\alpha}+[b^{-1}]_{\alpha})\tau_{\alpha \beta}
({\bf n}-{\bf e}_{\alpha}, {\bf t}).
\end{array}
\eeq
Besides, there are equations obtained from these by the changes
$\alpha \leftrightarrow \beta$, $a\leftrightarrow b$. We do not 
write them explicitly because in the dispersionless limit they
give the same results as equations (\ref{f12}), (\ref{f14}). 

The next choice of the variables in (\ref{f7}) is ${\bf n}'={\bf n}$,
${\bf t}-{\bf t}' =[a^{-1}]_{\mu}$, where $\mu \neq \alpha , \beta$
(and $\alpha \neq \beta$). The residue calculus gives the equation
\beq\label{f15}
\begin{array}{c}
\tau ({\bf n}, {\bf t})
\tau_{\alpha \beta}({\bf n}, {\bf t}+[a^{-1}]_{\mu})-
\tau ({\bf n}, {\bf t}+[a^{-1}]_{\mu})
\tau_{\alpha \beta}({\bf n}, {\bf t})
\\ \\
+\epsilon_{\alpha \beta}\epsilon_{\alpha \mu}\epsilon_{\beta \mu}a^{-1}
\Bigl (\tau_{\alpha \mu}({\bf n}, {\bf t})\tau_{\mu \beta}
({\bf n}, {\bf t}+[a^{-1}]_{\mu})-\tau_{\beta \mu}({\bf n}-{\bf e}_{\beta},
{\bf t})\tau_{\mu \alpha}({\bf n}+{\bf e}_{\alpha},
{\bf t}+[a^{-1}]_{\mu})\Bigr )=0.
\end{array}
\eeq

Now let us obtain equations with derivatives with 
respect to $t_{\alpha , 1}$.
Differentiate (\ref{f7}) with respect to $t_{\alpha , 1}$ and
put ${\bf n}'={\bf n}$,
${\bf t}-{\bf t}'=[a^{-1}]_{\alpha}+
[b^{-1}]_{\beta}$ after that. The residue calculus followed by some
simple transformations yields the equation
\beq\label{f16}
\begin{array}{c}
\displaystyle{
\frac{\tau ({\bf n}, {\bf t}+[a^{-1}]_{\alpha}+[b^{-1}]_{\beta})
\tau_{\alpha \beta}({\bf n}, {\bf t})}{\tau ({\bf n}, 
{\bf t}+[b^{-1}]_{\beta})\tau_{\alpha \beta}({\bf n}, {\bf t}+[a^{-1}]_{\alpha})}-a^{-2}
\frac{\tau ({\bf n}+{\bf e}_{\alpha}, {\bf t}+[a^{-1}]_{\alpha}+[b^{-1}]_{\beta})
\tau_{\alpha \beta}({\bf n}-{\bf e}_{\alpha}, {\bf t})}{\tau ({\bf n}, 
{\bf t}+[b^{-1}]_{\beta})\tau_{\alpha \beta}({\bf n}, {\bf t}+[a^{-1}]_{\alpha})}}
\\ \\
\displaystyle{
=1-a^{-1}\p_{t_{\alpha , 1}}
\log \frac{\tau_{\alpha \beta}
({\bf n}, {\bf t}+[a^{-1}]_{\alpha})}{\tau ({\bf n}, 
{\bf t}+[b^{-1}]_{\beta})}.}
\end{array}
\eeq
Another equation is obtained by putting 
${\bf n}'={\bf n}+{\bf e}_{\alpha}$,
${\bf t}-{\bf t}'=[a^{-1}]_{\alpha}+
[b^{-1}]_{\beta}$ in (\ref{f7}). The residue calculus yields:
\beq\label{f17}
\begin{array}{c}
\displaystyle{
\frac{\tau ({\bf n}+{\bf e}_{\alpha}, 
{\bf t}+[b^{-1}]_{\beta})\tau_{\alpha \beta}({\bf n}, {\bf t}+[a^{-1}]_{\alpha})}{\tau ({\bf n}+{\bf e}_{\alpha}, 
{\bf t}+[a^{-1}]_{\alpha}+[b^{-1}]_{\beta})
\tau_{\alpha \beta}({\bf n}, {\bf t})}-
a^{-2}
\frac{\tau ({\bf n}, {\bf t}+[b^{-1}]_{\beta})
\tau_{\alpha \beta}({\bf n}+{\bf e}_{\alpha}, 
{\bf t}+[a^{-1}]_{\alpha})}{\tau ({\bf n}+{\bf e}_{\alpha}, 
{\bf t}+[a^{-1}]_{\alpha}+[b^{-1}]_{\beta})
\tau_{\alpha \beta}({\bf n}, {\bf t})}}
\\ \\
\displaystyle{
=1-a^{-1}\p_{t_{\alpha , 1}}
\log \frac{\tau ({\bf n}+{\bf e}_{\alpha}, 
{\bf t}+[a^{-1}]_{\alpha}+[b^{-1}]_{\beta})}{\tau_{\alpha \beta}
({\bf n}, {\bf t})}.}
\end{array}
\eeq

Other two equations are obtained in a similar way 
from (\ref{f7}) at $\beta =\alpha$
(i.e. from (\ref{f9})) with the choices ${\bf t}-{\bf t}'=
[a^{-1}]_{\mu}+[b^{-1}]_{\mu}+[c^{-1}]_{\mu}$, ${\bf n}'={\bf n}+
{\bf e}_{\alpha}-{\bf e}_{\mu}$ and 
${\bf t}-{\bf t}'=
[a^{-1}]_{\mu}+[b^{-1}]_{\mu}+[c^{-1}]_{\mu}$, ${\bf n}'={\bf n}+
{\bf e}_{\alpha}-2{\bf e}_{\mu}$ ($\mu \neq \alpha$). The equations are:
\beq\label{f18}
\begin{array}{c}
(a-b)\tau_{\alpha \mu}({\bf n}, {\bf t}+[a^{-1}]_{\mu}+[b^{-1}]_{\mu})
\tau_{\alpha \mu}({\bf n}, {\bf t}+[c^{-1}]_{\mu})
\\ \\
+\, (b-c)\tau_{\alpha \mu}({\bf n}, {\bf t}+[b^{-1}]_{\mu}+[c^{-1}]_{\mu})
\tau_{\alpha \mu}({\bf n}, {\bf t}+[c^{-1}]_{\mu})
\\ \\
+\, (c-a)\tau_{\alpha \mu}({\bf n}, {\bf t}+[c^{-1}]_{\mu}+[a^{-1}]_{\mu})
\tau_{\alpha \mu}({\bf n}, {\bf t}+[b^{-1}]_{\mu})
\\ \\
\displaystyle{
+\, \frac{(a-b)(b-c)(c-a)}{(abc)^2}
\tau_{\alpha \mu}({\bf n}-{\bf e}_{\mu}, {\bf t})
\tau_{\alpha \mu}({\bf n}+{\bf e}_{\alpha}, {\bf t}+[a^{-1}]_{\mu}+[b^{-1}]_{\mu}+[c^{-1}]_{\mu}))=0,}
\end{array}
\eeq

\beq\label{f19}
\begin{array}{c}
(a-b)c^2\tau_{\alpha \mu}({\bf n}, {\bf t}+[a^{-1}]_{\mu}+[b^{-1}]_{\mu})
\tau_{\alpha \mu}({\bf n}-{\bf e}_{\mu}, {\bf t}+[c^{-1}]_{\mu})
\\ \\
+\, (b-c)a^2\tau_{\alpha \mu}({\bf n}, {\bf t}+[b^{-1}]_{\mu}+[c^{-1}]_{\mu})
\tau_{\alpha \mu}({\bf n}-{\bf e}_{\mu}, {\bf t}+[c^{-1}]_{\mu})
\\ \\
+\, (c-a)b^2\tau_{\alpha \mu}({\bf n}, {\bf t}+[c^{-1}]_{\mu}+[a^{-1}]_{\mu})
\tau_{\alpha \mu}({\bf n}-{\bf e}_{\mu}, {\bf t}+[b^{-1}]_{\mu})
\\ \\
\displaystyle{
+\, (a-b)(b-c)(c-a)
\tau_{\alpha \mu}({\bf n}-{\bf e}_{\mu}, {\bf t})
\tau_{\alpha \mu}({\bf n}, {\bf t}+[a^{-1}]_{\mu}+[b^{-1}]_{\mu}+[c^{-1}]_{\mu}))=0.}
\end{array}
\eeq

\section{The dispersionless limit}

In order to perform the dispersionless limit, one should introduce a small
parameter $\hbar$ and rescale the times ${\bf t}$ 
and variables ${\bf n}$ as
$t_{\alpha , k}\to t_{\alpha , k}/\hbar$, 
$n_{\alpha}\to t_{\alpha , 0}/\hbar$. Introduce a function 
$F({\bf t}_0, {\bf t}; \hbar )$ related to the tau-function by the
formula
\beq\label{d1}
\tau ({\bf t}_0/\hbar , {\bf t}/\hbar )=\exp \Bigl ( \frac{1}{\hbar^2}\,
F({\bf t}_0, {\bf t}; \hbar )\Bigr )
\eeq
and consider the limit $F=\lim\limits_{\hbar \to 0}
F({\bf t}_0, {\bf t}; \hbar )$. The function $F$ substitutes the 
tau-function in the dispersionless limit $\hbar \to 0$. It satisfies 
an infinite number of nonlinear differential equations which follow
from the bilinear equations for the tau-function. In the
dispersionless limit
the variables $t_{\alpha , 0}$ become continuous, and the
derivative $\p_{t_{\alpha , 0}}$ will be denoted as $\p_{\alpha}$
for short. Introduce also the differential operators 
\beq\label{d2}
D_{\alpha}(z)=\sum_{k\geq 1} \frac{z^{-k}}{k}\, \p_{t_{\alpha , k}},
\eeq
so that
$$
\tau ({\bf t}_0/\hbar , {\bf t}/\hbar +[z^{-1}]_{\alpha})
=\exp \Bigl ( \frac{1}{\hbar^2}
e^{\hbar D_{\alpha}(z)}
F({\bf t}_0, {\bf t}; \hbar )\Bigr ).
$$
As for the function $\tau_{\alpha \beta}$, we have by its definition:
\beq\label{d3}
\tau_{\alpha \beta}({\bf t}_0/\hbar , 
{\bf t}/\hbar )=\exp \Bigl ( \frac{1}{\hbar^2}\,
e^{\frac{\hbar}{2}\p_{\alpha}-
\frac{\hbar}{2}\p_{\beta}}F({\bf t}_0, {\bf t}; \hbar )\Bigr ).
\eeq

\subsection{The dispersionless limit of equations of the Hirota-Miwa 
type}

Let us rewrite the bilinear equations for the tau-function 
of the Hirota-Miwa type in terms of the function $F$ making them 
ready for the limit $\hbar \to 0$. We begin with equation (\ref{f10}).
In terms of the $F$-function it reads
$$
a^2\exp \left (\frac{1}{\hbar^2}\Bigl (e^{\hbar \p_{\alpha}
+\hbar D_{\alpha}(b)}+e^{\hbar D_{\alpha}(a)}-
e^{\hbar (\p_{\alpha} +D_{\alpha}(a)+D_{\alpha}(b))}-1\Bigr )F\right )
$$
$$
-b^2\exp \left (\frac{1}{\hbar^2}\Bigl (e^{\hbar \p_{\alpha}
+\hbar D_{\alpha}(a)}+e^{\hbar D_{\alpha}(b)}-
e^{\hbar (\p_{\alpha} +D_{\alpha}(a)+D_{\alpha}(b))}-1\Bigr )F\right )
$$
$$
=a^2-b^2 -(a-b)\hbar^{-1}\Bigl (e^{\hbar (\p_{\alpha} +D_{\alpha}(a)+D_{\alpha}(b))}-1\Bigr )F.
$$
In this form, the limit $\hbar \to 0$ is straightforward. The result is
\beq\label{f10a}
e^{-D_{\alpha}(a)D_{\alpha}(b)F}\Bigl (a^2 e^{-D_{\alpha}(a)\p_{\alpha}F}
-b^2 e^{-D_{\alpha}(b)\p_{\alpha}F}\Bigr )=a^2-b^2 -(a-b)
\Bigl ((\p_{\alpha}+D_{\alpha}(a)+D_{\alpha}(b))\p_{t_{\alpha , 1}}F
\Bigr ).
\eeq
In a similar way, the dispersionless limit of equation (\ref{f11}) reads
\beq\label{f11a}
e^{D_{\alpha}(a)D_{\alpha}(b)F}\Bigl (1-\frac{1}{a^2b^2}\,
e^{(\p_{\alpha}+D_{\alpha}(a)+D_{\alpha}(b))\p_{\alpha}F}\Bigr )=
1-\frac{1}{a-b}\, (D_{\alpha}(a)-D_{\alpha}(b))\p_{t_{\alpha , 1}}F.
\eeq

Equation (\ref{f111}) can be rewritten as
$$
\exp \left (\frac{1}{\hbar^2}\Bigl (e^{\hbar D_{\beta}(b)}+
e^{\hbar D_{\alpha}(a)}\Bigr )F\right )-
\exp \left (\frac{1}{\hbar^2}\Bigl (1+
e^{\hbar D_{\alpha}(a)+\hbar D_{\beta}(b)}\Bigr )F\right )
$$
$$
+\frac{1}{ab}\exp \left (\frac{1}{\hbar^2}\Bigl (
e^{\frac{\hbar}{2}\p_{\alpha}-\frac{\hbar}{2}\p_{\beta}+
\hbar D_{\alpha}(a)}+
e^{\frac{-\hbar}{2}\p_{\alpha}+\frac{\hbar}{2}\p_{\beta}+
\hbar D_{\beta}(b)}\Bigr )F\right )
$$
$$
-\frac{1}{ab}\exp \left (\frac{1}{\hbar^2}\Bigl (
e^{-\frac{\hbar}{2}\p_{\alpha}-\frac{\hbar}{2}\p_{\beta}}+
e^{\frac{\hbar}{2}\p_{\alpha}+\frac{\hbar}{2}\p_{\beta}+
\hbar D_{\alpha}(a)+
\hbar D_{\beta}(b)}\Bigr )F\right )=0.
$$
The limit $\hbar \to 0$ is straightforward:
\beq\label{f111a}
\begin{array}{c}
\displaystyle{
1-e^{D_{\alpha}(a)D_{\beta}(b)F}
+\frac{1}{ab}e^{\frac{1}{4}(\p_{\alpha}-\p_{\beta})^2F +
\frac{1}{2}(D_{\alpha}(a)-D_{\beta}(b))(\p_{\alpha}-\p_{\beta})F}}
\\ \\
\displaystyle{
-\frac{1}{ab}e^{\frac{1}{4}(\p_{\alpha}+\p_{\beta})^2F +
\frac{1}{2}(D_{\alpha}(a)-D_{\beta}(b))(\p_{\alpha}+\p_{\beta})F+
D_{\alpha}(a)D_{\beta}(b)F}}=0.
\end{array}
\eeq

Equation (\ref{f12}) can be rewritten as
$$
a\exp \left (\frac{1}{\hbar^2}\Bigl (e^{\hbar D_{\alpha}(a)}+
e^{\hbar D_{\alpha}(b)+\frac{\hbar}{2}\p_{\alpha}-
\frac{\hbar}{2}\p_{\beta}}\Bigr )F\right )
-b\exp \left (\frac{1}{\hbar^2}\Bigl (e^{\hbar D_{\alpha}(b)}+
e^{\hbar D_{\alpha}(a)+\frac{\hbar}{2}\p_{\alpha}-
\frac{\hbar}{2}\p_{\beta}}\Bigr )F\right )
$$
$$
=(a-b)\exp \left (\frac{1}{\hbar^2}\Bigl (1+
e^{\hbar D_{\alpha}(a)+\hbar D_{\alpha}(b)+
\frac{\hbar}{2}\p_{\alpha}-
\frac{\hbar}{2}\p_{\beta}}\Bigr )F\right )
$$
$$+
(a^{-1}-b^{-1})\exp \left (\frac{1}{\hbar^2}\Bigl (
e^{\hbar D_{\alpha}(a)+\hbar D_{\alpha}(b)+\hbar \p_{\alpha}}+
e^{-\frac{\hbar}{2}\p_{\alpha}-
\frac{\hbar}{2}\p_{\beta}}\Bigr )F\right ).
$$
The limit $\hbar \to 0$ yields:
\beq\label{f12a}
\begin{array}{c}
ae^{-\frac{1}{2}D_{\alpha}(a)(\p_{\alpha}-\p_{\beta})F}-
be^{-\frac{1}{2}D_{\alpha}(b)(\p_{\alpha}-\p_{\beta})F}
\\ \\
=(a-b)e^{D_{\alpha} (a)D_{\alpha}(b)F}\left (
1-(ab)^{-1}e^{\frac{1}{2}(\p_{\alpha}+D_{\alpha}(a)+D_{\alpha}(b))
(\p_{\alpha}+\p_{\beta})F}\right ).
\end{array}
\eeq
In a similar way, the dispersionless limit of equation (\ref{f14}) reads
\beq\label{f14a}
\begin{array}{c}
ae^{-\frac{1}{2}D_{\alpha}(a)(\p_{\alpha}+\p_{\beta})F}-
be^{-\frac{1}{2}D_{\alpha}(b)(\p_{\alpha}+\p_{\beta})F}
\\ \\
=(a-b)e^{D_{\alpha} (a)D_{\alpha}(b)F}\left (
1-(ab)^{-1}e^{\frac{1}{2}(\p_{\alpha}+D_{\alpha}(a)+D_{\alpha}(b))
(\p_{\alpha}-\p_{\beta})F}\right ).
\end{array}
\eeq

The limit of equation (\ref{f15}) is
\beq\label{f15a}
\begin{array}{c}
\displaystyle{
e^{-\frac{1}{4}\p_{\alpha}\p_{\beta}F}\Bigl (
1-e^{-\frac{1}{2}D_{\mu}(a)(\p_{\alpha}-\p_{\beta})F}\Bigr )}
\\ \\
\displaystyle{
+\epsilon_{\alpha \beta}\epsilon_{\alpha \mu}\epsilon_{\beta \mu}
a^{-1}\Bigl (e^{\frac{1}{4}(\p_{\mu}-\p_{\alpha}-\p_{\beta})\p_{\mu}F 
+\frac{1}{2}D_{\mu}(a)(\p_{\mu}-\p_{\alpha})F}-
e^{\frac{1}{4}(\p_{\mu}+\p_{\alpha}+\p_{\beta})\p_{\mu}F 
+\frac{1}{2}D_{\mu}(a)(\p_{\mu}+\p_{\beta})F}\Bigr )=0}.
\end{array}
\eeq

The dispersionless limits of equations (\ref{f16}), (\ref{f17})
can be found in a similar way. The result is:
\beq\label{f16a}
\begin{array}{c}
\displaystyle{
e^{D_{\alpha}(a)D_{\beta}(b)F}\Bigl (
ae^{-\frac{1}{2}D_{\alpha}(a)(\p_{\alpha}-\p_{\beta})F}-
a^{-1} e^{\frac{1}{2}(D_{\alpha}(a)+\p_{\alpha}+\p_{\beta})\p_{\alpha}F
+\frac{1}{2}D_{\alpha}(a)\p_{\beta}F+D_{\beta}(b)\p_{\alpha}F}\Bigr )
}
\\ \\
=a-\Bigl (\frac{1}{2}\p_{\alpha}+D_{\alpha}(a)-\frac{1}{2}\p_{\beta}
-D_{\beta}(b)\Bigr )\p_{t_{\alpha , 1}}F,
\end{array}
\eeq
\beq\label{f17a}
\begin{array}{c}
\displaystyle{
e^{-D_{\alpha}(a)D_{\beta}(b)F}\Bigl (
ae^{-\frac{1}{2}D_{\alpha}(a)(\p_{\alpha}+\p_{\beta})F}-
a^{-1} e^{\frac{1}{2}(D_{\alpha}(a)+\p_{\alpha}-\p_{\beta})\p_{\alpha}F
-\frac{1}{2}D_{\alpha}(a)\p_{\beta}F-D_{\beta}(b)\p_{\alpha}F}\Bigr )
}
\\ \\
=a-\Bigl (\frac{1}{2}\p_{\alpha}+D_{\alpha}(a)+\frac{1}{2}\p_{\beta}
+D_{\beta}(b)\Bigr )\p_{t_{\alpha , 1}}F.
\end{array}
\eeq
The dispersionless limits of equations (\ref{f18}), (\ref{f19}) are:
\beq\label{f18a}
\begin{array}{c}
(a-b)e^{D_{\mu}(a)D_{\mu}(b)F}+(b-c)e^{D_{\mu}(b)D_{\mu}(c)F}+
(c-a)e^{D_{\mu}(a)D_{\mu}(c)F}
\\ \\
\displaystyle{
+ \frac{(a\! -\! b)(b\! -\! c)(c\! -\! a)}{(abc)^2}\, 
e^{(\p_{\mu}+D_{\mu}(a)+
D_{\mu}(b)+D_{\mu}(c))\p_{\mu}F +D_{\mu}(a)D_{\mu}(b)F+
D_{\mu}(b)D_{\mu}(c)F+D_{\mu}(a)D_{\mu}(c)F}=0},
\end{array}
\eeq
\beq\label{f19a}
\begin{array}{c}
(a-b)c^2e^{D_{\mu}(a)D_{\mu}(b)F-D_{\mu}(c)\p_{\mu}F}
+(b-c)a^2e^{D_{\mu}(b)D_{\mu}(c)F-D_{\mu}(a)\p_{\mu}F}
\\ \\+
(c-a)b^2e^{D_{\mu}(a)D_{\mu}(c)F-D_{\mu}(b)\p_{\mu}F}
\\ \\
+(a\! -\! b)(b\! -\! c)(c\! -\! a)e^{
D_{\mu}(a)D_{\mu}(b)F+
D_{\mu}(b)D_{\mu}(c)F+D_{\mu}(a)D_{\mu}(c)F}=0.
\end{array}
\eeq

Let us rewrite these equations in a more suggestive form.
For this purpose, we introduce the notation
\beq\label{d4}
R_{\alpha}=e^{\frac{1}{4}\p_{\alpha}^2F}, \qquad 
R_{\alpha \beta}=R_{\beta \alpha}=e^{\frac{1}{4}\p_{\alpha}\p_{\beta}F}
\eeq
and the functions
\beq\label{d5}
\begin{array}{l}
\displaystyle{
w_{\alpha}(z)=ze^{-\frac{1}{2}D_{\alpha} (z)\p_{\alpha}F -
\frac{1}{4}\p_{\alpha}^2F},}
\\ \\
\displaystyle{
w_{\alpha \beta}(z)=e^{-\frac{1}{2}D_{\alpha} (z)\p_{\beta}F
-\frac{1}{4}\p_{\alpha}\p_{\beta}F}},
\end{array}
\eeq
\beq\label{d6}
\begin{array}{l}
p_{\alpha}(z)=z -(\frac{1}{2}\p_{\alpha}+D_{\alpha}(z))
\p_{t_{\alpha , 1}}F,
\\ \\
p_{\alpha \beta}(z)=-(\frac{1}{2}\p_{\alpha}+D_{\alpha}(z))
\p_{t_{\beta , 1}}F
\end{array}
\eeq
(note that $R_{\alpha \alpha}=R_{\alpha}$ but 
$w_{\alpha \alpha}(z)\neq w_{\alpha}(z)$, 
$p_{\alpha \alpha}(z)\neq p_{\alpha}(z)$).
In this notation, equations (\ref{f10a})--(\ref{f17a}) read:
\beq\label{f10b}
\begin{array}{l}
R_{\alpha}^2 e^{-D_{\alpha}(a)D_{\alpha}(b)F}
\Bigl (w_{\alpha}^2(a)-w_{\alpha}^2(b)\Bigr )=
(a-b)(p_{\alpha}(a)+p_{\alpha}(b)),
\\ \\
\displaystyle{
e^{D_{\alpha}(a)D_{\alpha}(b)F}\Bigl (1-w_{\alpha}^{-2}(a)
w_{\alpha}^{-2}(b)\Bigr )
=\frac{p_{\alpha}(a)-p_{\alpha}(b)}{a-b}},
\end{array}
\eeq

\beq\label{f111b}
1+\frac{w_{\alpha \beta}(a) w_{\beta \alpha}(b)}{w_{\alpha}(a)
w_{\beta}(b)}=e^{D_{\alpha}(a)D_{\beta}(b)F}
\left (1+\frac{1}{w_{\alpha}(a)
w_{\beta}(b)w_{\alpha \beta}(a) w_{\beta \alpha}(b)}\right ),
\eeq

\beq\label{f12b}
\begin{array}{l}
\displaystyle{
\frac{R_{\alpha}}{R_{\alpha \beta}}\left (
\frac{w_{\alpha}(a)}{w_{\alpha \beta}(a)}-
\frac{w_{\alpha}(b)}{w_{\alpha \beta}(b)}\right )=
(a-b)e^{D_{\alpha}(a)D_{\alpha}(b)F}\left (1-
\frac{1}{w_{\alpha}(a)w_{\alpha}(b)w_{\alpha \beta}(a)w_{\alpha \beta} (b)}
\right ),}
\\ \\
\displaystyle{
R_{\alpha}R_{\alpha \beta}\Bigl (
w_{\alpha}(a)w_{\alpha \beta}(a)-w_{\alpha}(b)w_{\alpha \beta}(b)\Bigr )=
(a-b)e^{D_{\alpha}(a)D_{\alpha}(b)F}\left (1-
\frac{w_{\alpha \beta}(a)
w_{\alpha \beta} (b)}{w_{\alpha}(a)w_{\alpha}(b)}\right ),}
\end{array}
\eeq

\beq\label{f14b}
w_{\mu}(a)\Bigl (R_{\mu \beta}w_{\mu \beta}(a)-
R_{\mu \alpha}w_{\mu \alpha}(a)\Bigr ) +\epsilon_{\alpha \beta}
\epsilon_{\alpha \mu}\epsilon_{\beta \mu}R_{\alpha \beta}\Bigl (
w_{\mu \alpha}(a)w_{\mu \beta}(a) -R_{\mu \alpha}R_{\mu \beta}\Bigr )=0,
\eeq

\beq\label{f16b}
\begin{array}{l}
\displaystyle{
\frac{R_{\alpha}}{R_{\alpha \beta}}\, e^{D_{\alpha}(a)D_{\beta}(b)F}
\left (w_{\alpha}(a)-\frac{1}{w_{\alpha}(a)w_{\beta \alpha}^2(b)}
\right )=w_{\alpha \beta}(a) (p_{\alpha}(a)-p_{\beta \alpha}(b))},
\\ \\
\displaystyle{
R_{\alpha}R_{\alpha \beta}\, 
e^{-D_{\alpha}(a)D_{\beta}(b)F}
\left (w_{\alpha}(a)-\frac{w_{\beta \alpha}^2(b)}{w_{\alpha}(a)}
\right )=w^{-1}_{\alpha \beta}(a) (p_{\alpha}(a)+p_{\beta \alpha}(b))}.
\end{array}
\eeq
In the next subsection we will show that the functions $w_{\alpha}(z)$,
$p_{\alpha}(z)$, $w_{\alpha \beta}(z)$, $p_{\beta \alpha}(z)$ are
connected by equation of an elliptic curve and represent equations
(\ref{f10b})--(\ref{f16b}) in the elliptic parametrization. 

\subsection{Elliptic parametrization}

Multiplying the two equations (\ref{f10b}), we get the relation
$$
R_{\alpha}^2\Bigl (w_{\alpha}^2(a)+w_{\alpha}^{-2}(a)\Bigr )-
p_{\alpha}^2(a)=R_{\alpha}^2\Bigl (w_{\alpha}^2(b)+
w_{\alpha}^{-2}(b)\Bigr )-
p_{\alpha}^2(b),
$$
from which it follows that the combination
$R_{\alpha}^2\Bigl (w_{\alpha}^2(z)+w_{\alpha}^{-2}(z)\Bigr )-
p_{\alpha}^2(z)=:V_{\alpha}$ does not depend on $z$. Tending $z\to 
\infty$, we find $V_{\alpha}=(\p_{t_{\alpha , 1}}\p_{\alpha}F)^2 +
2\p_{t_{\alpha ,1}}^2F-\p_{t_{\alpha, 2}}\p_{\alpha}F$. Therefore,
the functions $w_{\alpha}$, $p_{\alpha}$ are connected by equation 
of elliptic curve
\beq\label{e1}
R_{\alpha}^2\Bigl (w_{\alpha}^2(z)+w_{\alpha}^{-2}(z)\Bigr )=
p_{\alpha}^2(z)+V_{\alpha}.
\eeq
In a similar way, dividing equations (\ref{f12b}), we conclude that
$$
\frac{w_{\alpha}(a)}{w_{\alpha \beta}(a)}-
\frac{w_{\alpha \beta}(a)}{w_{\alpha}(a)}-
R^2_{\alpha \beta}\Bigl (w_{\alpha}(a)w_{\alpha \beta}(a)+
\frac{1}{w_{\alpha}(a)w_{\alpha \beta}(a)}\Bigr )=: V_{\alpha \beta}
$$
does not depend on $a$. The limit $a\to \infty$ yields:
\beq\label{e2}
V_{\alpha \beta}=e^{\frac{1}{4}(\p_{\beta}-\p_{\alpha})\p_{\alpha}F}
\p_{\beta}\p_{t_{\alpha ,1}}F.
\eeq
Therefore,
the functions $w_{\alpha}$, $w_{\alpha \beta}$ are connected by equation 
of elliptic curve
\beq\label{e3}
R_{\alpha \beta}^2\Bigl (w_{\alpha}^2(z)w_{\alpha \beta}^2(z)+1
\Bigr ) -\Bigl (w_{\alpha}^2(z)+ w_{\alpha \beta}^2(z)\Bigr )
+V_{\alpha \beta}w_{\alpha}(z)w_{\alpha \beta}(z)=0.
\eeq
Multiplying the two equations (\ref{f16b}), we conclude that the functions
$w_{\beta \alpha}(z)$ and $p_{\beta \alpha}(z)$ are connected by 
the same equation as (\ref{e1}):
\beq\label{e4}
R_{\alpha}^2\Bigl (w_{\beta \alpha}^2(z)+w_{\beta \alpha}^{-2}(z)\Bigr )=
p_{\beta \alpha}^2(z)+V_{\alpha}.
\eeq
In fact the three equations (\ref{e1}), (\ref{e3}), (\ref{e4}) define
the same elliptic curve. 

Our next goal is to uniformize this curve by elliptic functions. 
(In the one-component case this was done in \cite{AZ14}.)
To this end, we use the standard Jacobi
theta-functions $\theta_a (u)=\theta_a (u|\tau )$ ($a=1,2,3,4$).
Their definition and basic properties are listed in Appendix C.
The elliptic parametrization of the curve (\ref{e1}) is as follows
(see Appendix D):
\beq\label{E1}
w_{\alpha}(z)=\frac{\theta_4(u_{\alpha}
(z))}{\theta_1(u_{\alpha}(z))}\,, \qquad
p_{\alpha}(z)=\gamma_{\alpha} \, 
\theta_4^2(0)\, \frac{\theta_2(u_{\alpha}(z))\,
\theta_3(u_{\alpha}(z))}{\theta_1(u_{\alpha}(z))\, \theta_4(u_{\alpha}(z))}\,,
\eeq
where $u_{\alpha}(z)=u_{\alpha}(z, {\bf t})$ is some function 
of $z$ and $\gamma_{\alpha}$ is a $z$-independent
factor, and
\beq\label{E2}
R_{\alpha}=\gamma_{\alpha}\, \theta_2(0)\, \theta_3(0)\,, \qquad
V_{\alpha}=\gamma_{\alpha}^2 \Bigl ( \theta_2^4(0)+\theta_3^4(0)\Bigr ).
\eeq
Here $\gamma_{\alpha}$ is an arbitrary parameter but we will see that
it can not be put equal to a fixed number like 1 because it is a 
dynamical variable, as well as the modular parameter $\tau$:
$\gamma_{\alpha} =\gamma_{\alpha} ({\bf t})$, $\tau =\tau ({\bf t})$.
We normalize $u_{\alpha}(z)$ by the condition 
$u_{\alpha}(\infty )=0$, then the expansion around $\infty$ is
\beq\label{E3}
u_{\alpha}(z, {\bf t})=\frac{c_{\alpha ,1}({\bf t})}{z}+
\frac{c_{\alpha ,2}({\bf t})}{z^2}+\ldots .
\eeq
Tending $z\to \infty$ in (\ref{E1}) and using the identity 
(\ref{theta1prime}), we obtain that $\gamma_{\alpha}({\bf t})=
\pi c_{\alpha , 1}({\bf t})$.

Note that the function $w_{\alpha}(z)$ is the same in 
equations (\ref{e1}) and (\ref{e3}). The uniformization of the curve
(\ref{e3}) is as follows (see Appendix D):
\beq\label{E3a}
w_{\alpha}(z)=\frac{\theta_4(u_{\alpha}
(z))}{\theta_1(u_{\alpha}(z))}\,, \qquad
w_{\alpha \beta}(z)=\frac{\theta_4(u_{\alpha}
(z)+\eta_{\alpha \beta})}{\theta_1(u_{\alpha}(z)+\eta_{\alpha \beta})}\,,
\eeq
and
\beq\label{E4}
R_{\alpha \beta}=\frac{\theta_1(\eta_{\alpha 
\beta})}{\theta_4(\eta_{\alpha \beta})}, \qquad
V_{\alpha \beta}=2\frac{\theta_4^2(0)\theta_2(\eta_{\alpha \beta})
\theta_3(\eta_{\alpha \beta})}{\theta_2(0)
\theta_3(0)\theta_4^2(\eta_{\alpha \beta})}.
\eeq
Here $\eta_{\alpha \beta}=\eta_{\alpha \beta}({\bf t})$ is 
a $z$-independent constant depending on the times. 
Since $R_{\alpha \beta}=R_{\beta \alpha}$, it should hold
$$
\frac{\theta_1(\eta_{\alpha 
\beta})}{\theta_4(\eta_{\alpha \beta})}=
\frac{\theta_1(\eta_{\beta 
\alpha})}{\theta_4(\eta_{\beta \alpha})}.
$$
However, the assumption that $\eta_{\alpha \beta}=\eta_{\beta \alpha}$
is wrong, as we shall see below. Instead, the following relation holds:
\beq\label{E4a}
\eta_{\alpha \beta}+\eta_{\beta \alpha}=1.
\eeq
As for equation (\ref{e4}), its uniformization looks as follows:
\beq\label{E5}
w_{\beta \alpha}(z)=\frac{\theta_4(u_{\beta}
(z)+\eta_{\beta \alpha})}{\theta_1(u_{\beta}(z)+\eta_{\beta \alpha})}\,, \quad
p_{\beta \alpha}(z)=\gamma_{\alpha} \, 
\theta_4^2(0)\, \frac{\theta_2(u_{\beta}(z)+\eta_{\beta \alpha})\,
\theta_3(u_{\beta}(z)+\eta_{\beta \alpha})}{\theta_1(u_{\beta}(z)
+\eta_{\beta \alpha})\, \theta_4(u_{\beta}(z)+\eta_{\beta \alpha})}\,.
\eeq
From compatibility of these equations it follows that the modular
parameter $\tau$ is the same for all $\alpha$. 

In the elliptic parametrization equation (\ref{f111b}) acquires the form
$$
e^{D_{\alpha}(a)D_{\beta}(b)F}=
\left (\frac{\theta_4(u_{\alpha}(a))
\theta_4(u_{\beta}(b))}{\theta_1(u_{\alpha}(a))\theta_1(u_{\beta}(b))}-
\frac{\theta_4(u_{\alpha}(a)+\eta_{\alpha \beta})
\theta_4(u_{\beta}(b)-\eta_{\alpha \beta})}{\theta_1(u_{\alpha}(a)
+\eta_{\alpha \beta})\theta_1(u_{\beta}(b)-
\eta_{\alpha \beta})}\right )
$$
$$
\phantom{aaaaaaaaaaa}\times \left (\frac{\theta_4(u_{\alpha}(a))
\theta_4(u_{\beta}(b))}{\theta_1(u_{\alpha}(a))\theta_1(u_{\beta}(b))}-
\frac{\theta_1(u_{\alpha}(a)+\eta_{\alpha \beta})
\theta_1(u_{\beta}(b)-\eta_{\alpha \beta})}{\theta_4(u_{\alpha}(a)
+\eta_{\alpha \beta})\theta_4(u_{\beta}(b)-\eta_{\alpha \beta})}\right )^{-1},
$$
where we took into account (\ref{E4a}). Introducing the differential
operator 
\beq\label{E6}
\nabla_{\alpha}(z)=\frac{1}{2}\p_{\alpha}+D_{\alpha}(z)
\eeq
and using the identities for the theta functions given below, 
one can represent
this equation in the form
\beq\label{E7}
e^{\nabla_{\alpha}(a)\nabla_{\beta}(b)F}=
\frac{\theta_1(u_{\alpha}(a)-u_{\beta}(b)+
\eta_{\alpha \beta})}{\theta_4(u_{\alpha}(a)-u_{\beta}(b)+
\eta_{\alpha \beta})}.
\eeq
We need the following identities:
\beq\label{id1}
\begin{array}{c}
\theta_4(x)\theta_4(y)\theta_1(x+\eta )\theta_1(y-\eta )
-\theta_1(x)\theta_1(y)\theta_4(x+\eta )\theta_4(y-\eta )
\\ \\
=-\theta_4(0)\theta_1(\eta )\theta_1(x-y+\eta )\theta_4(x+y),
\end{array}
\eeq
\beq\label{id2}
\begin{array}{c}
\theta_4(x)\theta_4(y)\theta_4(x+\eta )\theta_4(y-\eta )
-\theta_1(x)\theta_1(y)\theta_1(x+\eta )\theta_1(y-\eta )
\\ \\
=\theta_4(0)\theta_4(\eta )\theta_4(x-y+\eta )\theta_4(x+y).
\end{array}
\eeq
In order to prove (\ref{id1}) we first notice that the three terms,
as functions of $x$, 
have the same monodromy properties under shifts by the periods and
the both sides equal zero at $x=y-\eta$ and $x=-y+\frac{\tau}{2}$. 
(Here we use the transformation properties of the theta-functions
given in Appendix B.)
Because the both sides are theta-functions of the second order, it is
enough to establish the equality at a third point; it is convenient 
to take $x=0$. The proof of (\ref{id2}) is similar. 
The same equation (\ref{E7}) 
is obtained from the elliptic parametrization of
equations (\ref{f16b}). 

In a similar way, equations (\ref{f10b}), (\ref{f12b}) 
in the elliptic parametrization are converted into the equation
\beq\label{f8}
(a^{-1}-b^{-1})e^{\nabla_{\alpha}(a)\nabla_{\alpha}(b)F}=
\frac{\theta_1(u_{\alpha}(a)-u_{\alpha}(b))}{\theta_4(u_{\alpha}(a)
-u_{\alpha}(b))}.
\eeq

Finally, equation (\ref{f14b}) in the elliptic parametrization 
acquires the form
$$
\epsilon_{\alpha \beta}\frac{\theta_1
(\eta_{\alpha \beta})}{\theta_4
(\eta_{\alpha \beta})}=\epsilon_{\alpha \mu}\epsilon_{\beta \mu}
\frac{\theta_1
(\eta_{\mu \alpha}-\eta_{\mu \beta})}{\theta_4
(\eta_{\mu \alpha}-\eta_{\mu \beta})}
$$
which imposes constraints on the quantities $\eta_{\alpha \beta}$.
The solution is
\beq\label{E9}
\eta_{\alpha \beta}=\eta_{\alpha}-\eta_{\beta}+\frac{1}{2}
(\epsilon_{\alpha \beta}+1)
\eeq
with some $\eta_{\alpha}$'s. This implies the already mentioned relation
$\eta_{\alpha \beta}+\eta_{\beta \alpha}=1$. 

Equations (\ref{f18a}), (\ref{f19a}) can be rewritten as
\beq\label{f18b}
\begin{array}{c}
(a^{-1}-b^{-1})w_{\mu}(a)w_{\mu}(b)e^{\nabla_{\mu}(a)\nabla_{\mu}(b)F}
+(b^{-1}-c^{-1})w_{\mu}(b)w_{\mu}(c)e^{\nabla_{\mu}(b)\nabla_{\mu}(c)F}
\\ \\
+(c^{-1}-a^{-1})w_{\mu}(c)w_{\mu}(a)e^{\nabla_{\mu}(c)\nabla_{\mu}(a)F}
\\ \\
\displaystyle{+(a^{-1}-b^{-1})(b^{-1}-c^{-1})(c^{-1}-a^{-1})
e^{\nabla_{\mu}(a)\nabla_{\mu}(b)F+\nabla_{\mu}(b)\nabla_{\mu}(c)F
+\nabla_{\mu}(c)\nabla_{\mu}(a)F}=0, }
\end{array}
\eeq

\beq\label{f19b}
\begin{array}{c}
(a^{-1}-b^{-1})w_{\mu}^{-1}(a)w_{\mu}^{-1}(b)
e^{\nabla_{\mu}(a)\nabla_{\mu}(b)F}
+(b^{-1}-c^{-1})w_{\mu}^{-1}(b)w_{\mu}^{-1}(c)
e^{\nabla_{\mu}(b)\nabla_{\mu}(c)F}
\\ \\
+(c^{-1}-a^{-1})w_{\mu}^{-1}(c)w_{\mu}^{-1}(a)
e^{\nabla_{\mu}(c)\nabla_{\mu}(a)F}
\\ \\
\displaystyle{+(a^{-1}-b^{-1})(b^{-1}-c^{-1})(c^{-1}-a^{-1})
e^{\nabla_{\mu}(a)\nabla_{\mu}(b)F+\nabla_{\mu}(b)\nabla_{\mu}(c)F
+\nabla_{\mu}(c)\nabla_{\mu}(a)F}=0. }
\end{array}
\eeq
In fact these equations follow from (\ref{f8}), as it follows from
the identity
\beq\label{id1a}
\begin{array}{c}
\displaystyle{
\frac{\theta_1(x-y)\theta_4(x)\theta_4(y)}{\theta_4(x-y)
\theta_1(x)\theta_1(y)}+
\frac{\theta_1(y-z)\theta_4(y)\theta_4(z)}{\theta_4(y-z)
\theta_1(y)\theta_1(z)}+
\frac{\theta_1(z-x)\theta_4(z)\theta_4(x)}{\theta_4(z-x)
\theta_1(z)\theta_1(x)}}
\\ \\
+\, \displaystyle{\frac{\theta_1(x-y)\theta_1(y-z)
\theta_1(z-x)}{\theta_4(x-y)\theta_4(y-z)
\theta_4(z-x)}=0}.
\end{array}
\eeq
The proof of this identity is standard. The left hans side is an 
elliptic function of $x$ with possible simple poles at $0$, 
$y+\frac{\tau}{2}$, $z+\frac{\tau}{2}$. It is not difficult to see that
the residues at these poles vanish. Therefore, the left hand side does not
depend on $x$. Taking $x=y$, we conclude that it equals zero. 

The final result is as follows. Let us redefine the functions
$u_{\alpha}(z)$ as $u_{\alpha}(z)\to u_{\alpha}(z)+\eta_{\alpha}$.
In other words, we introduce the functions
\beq\label{v}
v_{\alpha}(z)=u_{\alpha}(z)+\eta_{\alpha}
\eeq
such that $v_{\alpha}(\infty )=\eta_{\alpha}$.
Then all equations of the hierarchy are encoded in the single equation
\beq\label{main}
\epsilon_{\beta \alpha}(a^{-1}-b^{-1})^{\delta_{\alpha \beta}}e^{\nabla_{\alpha}(a)\nabla_{\beta}(b)F}
=\frac{\theta_1 (v_{\alpha}(a)-
v_{\beta}(b))}{\theta_4 (v_{\alpha}(a)-v_{\beta}(b))}.
\eeq
The meaning of this equation is that the general second order derivatives
of the function $F$ are expressed through some special second order
derivatives. 

Finally, by applying $\nabla_{\gamma}(c)$ to the logarithm of both
sides of (\ref{main}), this equation can be rewritten as
\beq\label{main1}
\begin{array}{c}
\displaystyle{
\nabla_{\alpha}(a)\log \frac{\theta_1 (v_{\beta}(b)-
v_{\gamma}(c))}{\theta_4 (v_{\beta}(b)-v_{\gamma}(c))}=
\nabla_{\beta}(b)\log \frac{\theta_1 (v_{\gamma}(c)-
v_{\alpha}(a))}{\theta_4 (v_{\gamma}(c)-v_{\alpha}(a))}}
\\ \\
\displaystyle{
=\nabla_{\gamma}(c)\log \frac{\theta_1 (v_{\alpha}(a)-
v_{\beta}(b))}{\theta_4 (v_{\alpha}(a)-v_{\beta}(b))}.}
\end{array}
\eeq
This symmetry is a manifistation of integrability of the dispersionless
multicomponent DKP hierarchy.

\section{Concluding remarks}

We have introduced the multicomponent DKP hierarchy by deriving the
basic integral bilinear equation for its tau-function in the framework
of the free fermions approach. We also obtained the various bilinear
equations of the Hirota-Miwa type as its corollaries. Next we have
obtained the hierarchy of nonlinear differential equations for
the dispersionless limit of the logarithm of the tau-function. 
We have shown that there is an elliptic curve naturally built in the 
structure of the hierarchy. This curve can be uniformized in terms of
elliptic functions. In the elliptic parametrization, the equations
of the hierarchy acquire an especially nice form being encoded in 
a single equation. 

When the elliptic modulus tends to infinity, the curve becomes
singular (a rational curve) and the elliptic functions become 
trigonometric ones. As is shown in the recent paper \cite{Z24},
in this case one deals with the multicomponent KP hierarchy
instead of DKP. 

A natural generalization of this work would be introducing the 
multicomponent Pfaff-Toda hierarchy and considering its dispersionless
limit in the elliptic parametrization. Also, the problem for future research
is to characterize the finite dimensional reductions of the dispersionless
hierarchy.

\section*{Appendix A: proof of the bilinear identity}
\addcontentsline{toc}{section}{Appendix A: proof of the bilinear identity}
\def\theequation{A\arabic{equation}}
\setcounter{equation}{0}

In this appendix we prove the fermionic bilinear identity (\ref{f2}).

First of all,
let us focus on the adjoint action of the general bilinear fermionic
element
$$
X=\sum_{\alpha , \beta}
\sum_{j,k}\Bigl ( A_{jk}^{(\alpha \beta )}
\psi^{(\alpha )}_{j}\psi^{*(\beta )}_{k}+
B_{jk}^{(\alpha \beta )}
\psi^{(\alpha )}_{j}\psi^{(\beta )}_{k}+
C_{jk}^{(\alpha \beta )}
\psi^{*(\alpha )}_{j}\psi^{*(\beta )}_{k}\Bigr ),
$$
where $A_{jk}^{(\alpha \beta )}$, 
$B_{jk}^{(\alpha \beta )}$, $C_{jk}^{(\alpha \beta )}$ 
are some infinite matrices. 
For simplicity of the notation, in what follows 
we omit the sum sign implying it for 
any repeated indices.
We will also use a form with matrix multiplication applied,
for example:
$$
     U^{(\mu \nu)}_{ij}\psi^{(\nu)}_{j} = U \psi ,
     \qquad
     \psi^{(\mu)}_{i}U^{(\mu \nu)}_{ij} = \psi U.
$$
Consider the adjoint action $\mbox{ad}_X(U \psi)$ of 
$X$ on some linear combination $U \psi$:
$$
    [X, U\psi] 
    =
    [X, U^{(\mu \nu)}_{pq}\psi^{(\nu)}_{q}]
    =
    A^{(\alpha \beta)}_{ij}U^{(\mu \nu)}_{pq}
    \psi^{(\alpha)}_{i}\delta_{jq}\delta^{\beta \nu}
    +
    C^{(\alpha \beta)}_{ij}U^{(\mu \nu)}_{pq}
    \bigl(
    \psi^{*(\alpha)}_{i}\delta_{jq}\delta^{\beta\nu}
    -
    \psi^{*(\beta)}_{j}\delta_{qi}\delta^{\alpha\nu}
    \bigl)
$$ 
$$=
    (U A^T)^{\mu \alpha }_{pi}\psi^{(\alpha)}_{i}
    +
    \bigl(U (C^T - C)\bigl)^{\mu \alpha}_{pi}
    \psi^{*(\alpha)}_{i}
    =
    (U A^T) \psi + U(C^T - C)\psi^{*},
$$
where ${(...)}^T$ means the transposed matrix. We see that the result 
is linear in $\psi^{(\alpha)}_{i}$
and $\psi^{*(\alpha)}_{i}$.
The same is true for the following actions:
$$
\begin{array}{lll}
\mbox{ad}_{X}(V\psi^{*})
    &=
    [X, V\psi^{*}] 
    =
    V(-A)\psi^{*} + V(B^T - B)\psi ,
    \\ &&\\
    \mbox{ad}_{X}(\psi U)
    &=
    [X, \psi U] 
    =
    \psi AU + \psi^* (C- C^T)V,
    \\ && \\
    \mbox{ad}_{X}(\psi^{*}V)
    &=
    [X, \psi^{*}V] 
    = 
    \psi^{*}(-A^T)V + \psi (B - B^T)V.
\end{array}
$$
These actions can not be written in terms of $\psi$ and $\psi^*$
separately, and the way to unify them is to consider the action
as multiplication by $2\times2$ block matrix:
$$
    \mbox{ad}_X
    \left (\begin{array}{l}
    \!\! U\psi \!\! 
    \\
    \!\! V\psi^* \!\!
    \end{array} \right )
    =
    \left (\begin{array}{l}
    UA^T\psi
    +
    U(C^T \! -\!  C)\psi^* 
    \\
    -VA\psi^*
    +
    V(B^T \! -\!  B)\psi
    \end{array}\right )
    =
    \left (\begin{array}{ll}
    U & 0 
    \\
    0 & V
    \end{array}\right )
    \left (\begin{array}{cc}
    A^T & C^T \! -\!  C 
    \\
    B^T \! -\!  B & -A
    \end{array}\right )
    \left (\begin{array}{l}
    \psi 
    \\
    \psi^*
    \end{array}\right ).
$$
In a similar way, we have:
$$
\mbox{ad}_X
    \Bigl (\begin{array}{ll}
    \psi^*V
    ,&
    \psi U
    \end{array} \Bigr )
    =
    \Bigl (\begin{array}{ll}
    \psi^*
    ,&
    \psi
    \end{array} \Bigr )
    \left (\begin{array}{cc}
    -A^T & C - C^T
    \\
    B - B^T & A
    \end{array}\right )
    \left (\begin{array}{cc}
    V & 0 
    \\
    0 & U
    \end{array}\right ).
$$
Therefore, applying $\mbox{ad}_X$ $n$ times
we get:
$$
    \mbox{ad}^n_X
    \left (\begin{array}{l}
    \psi 
    \\
    \psi^*
    \end{array} \right )
    =
    \left (\begin{array}{cc}
    A^T & C^T - C 
    \\
    B^T - B & -A
    \end{array}
    \right )^n
    \left (\begin{array}{l}
    \psi 
    \\
    \psi^*
    \end{array} \right )
    =
    R^{n}
    \left (\begin{array}{l}
    \psi 
    \\
    \psi^*
    \end{array} \right ),
$$
$$
    \mbox{ad}^n_X
    \Bigl (\begin{array}{ll}
    \psi^*
    ,&
    \psi
    \end{array}\Bigr )
    =
    \Bigl (\begin{array}{ll}
    \psi^*
    ,&
    \psi
    \end{array}\Bigr )
    \left (\begin{array}{cc}
    -A^T & C - C^T
    \\
    B - B^T & A
    \end{array} \right )^n
    =
    \Bigl (\begin{array}{ll}
    \psi^*
    ,&
    \psi
    \end{array}\Bigr )
    (-R)^{n},
$$
where the rotation matrix $R$ has the form
$$
R =
    \left ( \begin{array}{cc}
    A^T & C^T - C 
    \\
    B^T - B & -A
    \end{array}\right ).
$$
Now we can find the adjoint action $e^X \phi e^{-X}$ using the formula
$$
e^{X}\phi e^{-X} 
    =
    \phi + [X,\phi ] +
    \frac{1}{2!}[X,[X,\phi ]] + 
    \frac{1}{3!}[X,[X,[X,\phi ]]] + 
    \dots
$$
$$ =
    \phi + \mbox{ad}_X(\phi ) 
    + \frac{1}{2!}\mbox{ad}^2_X(\phi ) 
    + \frac{1}{3!}\mbox{ad}^3_X(\phi )
    + \dots
$$
Therefore, we have:
\begin{equation}\label{adj}
    e^X
    \left (\begin{array}{l}
    \psi 
    \\
    \psi^*
    \end{array} \right )
    e^{-X}
    =
    e^{R}
    \left (\begin{array}{l}
    \psi 
    \\
    \psi^*
    \end{array} \right ),
    \qquad
    e^X
    \Bigl (\begin{array}{ll}
    \psi^*
    ,&
    \psi
    \end{array}\Bigr )
    e^{-X}
    =
    \Bigl (\begin{array}{ll}
    \psi^*
    ,&
    \psi
    \end{array}\Bigr )
    e^{-R}.
    \end{equation}
    
    With the use of formulas (\ref{adj}) 
we can prove the fermionic bilinear relation for the group
element $g = \exp(X)$:
\begin{equation}\label{bil-rel-fermions}
    \sum_{\gamma = 1}^{N}
    \sum_{n \in \z}
    \bigl( 
    g\psi^{(\gamma)}_n \otimes g\psi^{*(\gamma)}_n 
    +
    g\psi^{*(\gamma)}_n \otimes g\psi^{(\gamma))}_n 
    \bigl)
    =
    \sum_{\gamma = 1}^{N}
    \sum_{n \in \z}
    \bigl( 
    \psi^{(\gamma)}_n g \otimes \psi^{*(\gamma)}_n g
    +
    \psi^{*(\gamma)}_n g \otimes \psi^{(\gamma)}_n g
    \bigl).
\end{equation}
For the proof consider the tensor product
$$
    \left (\begin{array}{l}
    \psi 
    \\
    \psi^*
    \end{array}\right )
    \otimes
    \Bigl (\begin{array}{ll}
        \psi^* , &
        \psi
    \end{array}\Bigr )
    =
    \left (\begin{array}{cc}
        \psi \otimes \psi^* & \psi \otimes \psi 
        \\
        \psi^* \otimes \psi^* & \psi^* \otimes \psi
    \end{array} \right ).
$$
There are three matrix structures inside it: $2 \times 2$ block matrix, 
$N \times N$ with respect to Greek indices, and $\infty \times \infty$ 
with respect to Latin indices. Applying the trace operation
with respect to all of them yields:
$$
    \mbox{tr}_{N \times N}
    \mbox{tr}_{\infty \times \infty}
    \mbox{tr}_{2 \times 2}
    \left ( \begin{array}{l}
    \psi 
    \\
    \psi^*
    \end{array}\right )
    \otimes
    \Bigl (\begin{array}{ll}
        \psi^* 
        ,&
        \psi
    \end{array}\Bigr )
    =
    \mbox{tr}_{N \times N}
    \mbox{tr}_{\infty \times \infty}
    \bigl(
    \psi \otimes \psi^* + \psi^* \otimes \psi
    \bigl)
    $$
    $$
    = 
    \mbox{tr}_{N \times N}
    \sum_{n \in \z}
    \bigl( 
    \psi_n \otimes \psi^{*}_n
    +
    \psi^{*}_n \otimes \psi_n
    \bigl)
    =
    \sum_{\gamma = 1}^{N}
    \sum_{n \in \z}
    \bigl( 
    \psi^{(\gamma)}_n \otimes \psi^{*(\gamma)}_n
    +
    \psi^{*(\gamma)}_n \otimes \psi^{(\gamma)}_n
    \bigl).
$$
The cyclic property of the trace
allows one to cancel opposite rotations, when one applies the 
adjoint action of 
$g = e^X$ on both row and column:
$$
    \mbox{tr}\left (
    g
    \left ( \begin{array}{l}
    \psi 
    \\
    \psi^*
    \end{array}\right )
    g^{-1}
    \otimes
    g
    \Bigl (\begin{array}{ll}
        \psi^* 
        ,&
        \psi
    \end{array}\Bigr )
    g^{-1}
    \right )
    =
    \mbox{tr}\left (
    e^{R}
    \left (\begin{array}{l}
    \psi 
    \\
    \psi^*
    \end{array}\right )
    \otimes
    \Bigl (\begin{array}{ll}
        \psi^* 
        ,&
        \psi
    \end{array}\Bigr )
    e^{-R}
    \right )
$$
$$   =
    \mbox{tr}
    \left (
    e^{-R}e^{R}
    \left (\begin{array}{l}
    \psi 
    \\
    \psi^*
    \end{array}\right )
    \otimes
    \Bigl ( \begin{array}{ll}
        \psi^* 
        ,&
        \psi
    \end{array}\Bigr )
    \right )
    =
    \mbox{tr}
    \left (
    \left ( \begin{array}{l}
    \psi 
    \\
    \psi^*
    \end{array}\right )
    \otimes
    \Bigl (\begin{array}{ll}
        \psi^* 
        ,&
        \psi
    \end{array}\Bigr )
    \right ).
$$
This proves the desired bilinear identity (\ref{f2}).

\section*{Appendix B: proof of the bosonization rules}
\addcontentsline{toc}{section}{Appendix B: proof of the bosonization rules}
\def\theequation{B\arabic{equation}}
\setcounter{equation}{0}

Let
$$
J_k^{(\alpha )}=\sum_{j\in \z}\normord \psi_{j}^{(\alpha )}
\psi_{j+k}^{*(\alpha )}\normord
$$
be modes of the current operator; their commutator is
$$
[J_k^{(\alpha )}, J_l^{(\beta )}]=k\delta_{\alpha \beta}\delta_{k+l,0}.
$$
We set
\beq\label{B101}
J_{\pm}^{(\alpha )}({\bf t})=\sum_{k\geq 1}t_{\alpha ,k}
J_{\pm k}^{(\alpha )}, \qquad
J_{\pm}({\bf t})=\sum_{\alpha}J_{\pm}^{(\alpha )}({\bf t}).
\eeq
These operators commute as
$$
[J_{+}^{(\alpha )}({\bf t}), J_{-}^{(\beta )}(\bar {\bf t})]=
\delta_{\alpha \beta}\sum_{k\geq 1}kt_{\alpha , k}\bar t_{\alpha , k}
$$
and
\beq\label{B102}
e^{J_{+}({\bf t})} e^{J_{-}(\bar {\bf t})}=
\exp \Bigl (\sum_{\alpha}\sum_{k\geq 1}kt_{\alpha ,k}
\bar t_{\alpha ,k}\Bigr )\, e^{J_{-}({\bf t})} e^{J_{+}(\bar {\bf t})}.
\eeq
Here $\bar {\bf t}$ is another set of variables independent of ${\bf t}$.
Note that
\beq\label{B103}
J_{+}({\bf t})\left |{\bf n}\rbr =\lbr {\bf n}\right |J_{-}({\bf t})=0.
\eeq
The commutation relations of the currents with the fermionic operators
are:
\beq\label{B104}
\begin{array}{l}
e^{J_{+}({\bf t})}\psi^{(\gamma )}(z)=e^{\xi ({\bf t}_{\gamma}, z)}
\psi^{(\gamma )}(z)e^{J_{+}({\bf t})},
\\ \\
e^{J_{+}({\bf t})}\psi^{*(\gamma )}(z)=e^{-\xi ({\bf t}_{\gamma}, z)}
\psi^{*(\gamma )}(z)e^{J_{+}({\bf t})},
\\ \\
e^{J_{-}({\bf t})}\psi^{(\gamma )}(z)=e^{\xi ({\bf t}_{\gamma}, z^{-1})}
\psi^{(\gamma )}(z)e^{J_{-}({\bf t})},
\\ \\
e^{J_{-}({\bf t})}\psi^{*(\gamma )}(z)=e^{-\xi ({\bf t}_{\gamma}, z^{-1})}
\psi^{*(\gamma )}(z)e^{J_{-}({\bf t})}.
\end{array}
\eeq

We will prove the bosonization rules in the form
\beq\label{B105}
\lbr {\bf n}\right |\psi^{(\gamma )}(z)e^{J_+({\bf t})}=
\epsilon_{\gamma }({\bf n})z^{n_{\gamma}-1}\lbr {\bf n}-{\bf e}_{\gamma}
\right | e^{J_{+}({\bf t}-[z^{-1}]_{\gamma})},
\eeq
\beq\label{B106}
\lbr {\bf n}\right |\psi^{*(\gamma )}(z)e^{J_+({\bf t})}=
\epsilon_{\gamma }({\bf n})z^{-n_{\gamma}}\lbr {\bf n}+{\bf e}_{\gamma}
\right | e^{J_{+}({\bf t}+[z^{-1}]_{\gamma})}.
\eeq
Let us prove (\ref{B105}).
It is enough to check that the both sides coincide by taking the scalar
product with the state $e^{J_-(\bar {\bf t})}\left |{\bf n}-{\bf e}_{\gamma}
\rbr$ for all $\bar {\bf t}$ 
which is known to be the generating function for basis 
states in the fermionic Hilbert space. Using (\ref{B104}), (\ref{B103})
and (\ref{B102}), we see that this is indeed the case. The proof of
(\ref{B106}) is similar.

\section*{Appendix C: theta-functions}
\addcontentsline{toc}{section}{Appendix C: theta-functions}
\def\theequation{C\arabic{equation}}
\setcounter{equation}{0}

The Jacobi's theta-functions $\theta_a (u)=
\theta_a (u|\tau )$, $a=1,2,3,4$, are defined by the formulas
\beq\label{Bp1}
\begin{array}{l}
\theta _1(u)=-\displaystyle{\sum _{k\in \z}}
\exp \left (
\pi i \tau (k+\frac{1}{2})^2 +2\pi i
(u+\frac{1}{2})(k+\frac{1}{2})\right ),
\\
\theta _2(u)=\displaystyle{\sum _{k\in \z}}
\exp \left (
\pi i \tau (k+\frac{1}{2})^2 +2\pi i
u(k+\frac{1}{2})\right ),
\\
\theta _3(u)=\displaystyle{\sum _{k\in \z}}
\exp \left (
\pi i \tau k^2 +2\pi i u k \right ),
\\
\theta _4(u)=\displaystyle{\sum _{k\in \z}}
\exp \left (
\pi i \tau k^2 +2\pi i
(u+\frac{1}{2})k\right ),
\end{array}
\eeq where $\tau$ is a complex parameter (the modular parameter) 
such that ${\rm Im}\, \tau >0$. The function 
$\theta_1(u)$ is odd, the other three functions are even.
The infinite product representation for the $\theta_1(u)$ reads: 
\beq
\label{infprod}
\theta_1(u|\tau)=2q^{\frac{1}{4}}\sin\pi u
\prod_{n=1}^\infty(1-q^{2n})(1-q^{2n}e^{2\pi i u})(1-q^{2n}e^{-2\pi i u}),
\eeq 
where $q=e^{\pi i \tau}$.
We also mention the identity
\beq\label{theta1prime}
\theta_1'(0)=\pi \theta_2(0) \theta_3(0) \theta_4(0).
\eeq

Here we list the transformation properties of the theta functions.

\smallskip

\noindent
Shifts by periods:
\beq\label{p}
\begin{array}{l}
\theta_1(u+1)=-\theta_1(u),\\
\theta_2(u+1)=-\theta_2(u),\\
\theta_3(u+1)=\theta_3(u),\\
\theta_4(u+1)=\theta_4(u).
\end{array}\hspace{2cm}
\begin{array}{l}
\theta_1(u+\tau)=-e^{-\pi i(2u+\tau)}\theta_1(u),\\
\theta_2(u+\tau)=e^{-\pi i(2u+\tau)}\theta_2(u),\\
\theta_3(u+\tau)=e^{-\pi i(2u+\tau)}\theta_3(u),\\
\theta_4(u+\tau)=-e^{-\pi i(2u+\tau)}\theta_4(u).
\end{array}\vspace{0.3cm}
\eeq
Shifts by half-periods:
\beq\label{hp}
\begin{array}{l}
\theta_1(u+{\textstyle\frac{1}{2}})=\theta_2(u),\\
\theta_2(u+{\textstyle\frac{1}{2}})=-\theta_1(u),\\
\theta_3(u+{\textstyle\frac{1}{2}})=\theta_4(u),\\
\theta_4(u+{\textstyle\frac{1}{2}})=\theta_3(u).
\end{array}\hspace{2cm}
\begin{array}{l}
\theta_1(u+{\textstyle\frac{\tau}{2}})=
ie^{-\pi i(u+\tau/4)}\theta_4(u),\\
\theta_2(u+{\textstyle\frac{\tau}{2}})=
e^{-\pi i(u+\tau/4)}\theta_3(u),\\
\theta_3(u+{\textstyle\frac{\tau}{2}})=
e^{-\pi i(u+\tau/4)}\theta_2(u),\\
\theta_4(u+{\textstyle\frac{\tau}{2}})=i
e^{-\pi i(u+\tau/4)}\theta_1(u).
\end{array}\vspace{0.3cm}
\eeq

For a more detailed account of properties of the theta-functions
see \cite{KZ15}. 

\section*{Appendix D: uniformization of elliptic curves}
\addcontentsline{toc}{section}{Appendix D: uniformization of elliptic curves}
\def\theequation{D\arabic{equation}}
\setcounter{equation}{0}

The equation $P(x,y)=0$, where $P(x,y)$ is a bi-quadratic polynomial
in the variables $x,y$ defines an elliptic curve. This curve can be
uniformized by elliptic functions $x=x(u)$, $y=y(u)$ of a complex 
variable $u$
such that the equation $P(x(u),y(u))=0$ becomes an identity. 

Our first example is the curve
\beq\label{C1}
y^2 -R^2 (x+x^{-1})+V=0.
\eeq
The uniformization is as follows:
\beq\label{C2}
x=\frac{\theta_4^2(u)}{\theta_1^2(u)}, \qquad
y=\theta_4^2(0)\frac{\theta_2(u)\theta_3(u)}{\theta_1(u)
\theta_4(u)}
\eeq
and
\beq\label{C3}
R=\theta_2(0)\theta_3(0), \qquad V=\theta_2^4(0)+\theta_3^4(0).
\eeq
Indeed, these functions satisfy the identity
\beq\label{C4}
\theta_4^4(0)\frac{\theta_2^2(u)\theta_3^2(u)}{\theta_1^2(u)
\theta_4^2(u)}-\theta_2^2(0)\theta_3^2(0)\left (
\frac{\theta_4^2(u)}{\theta_1^2(u)}+\frac{\theta_1^2(u)}{\theta_4^2(u)}
\right )+\theta_2^4(0)+\theta_3^4(0)=0.
\eeq
For the proof we note that the left hand side is an even elliptic
function of $u$ having possible poles at $u=0$ and $u=\frac{\tau}{2}$.
However, the expansion around these points shows that the singular terms
cancel and the function is regular everywhere. This means that it is
a constant. To find the constant one can substitute any value of $u$.
It is convenient to take $u=\frac{1}{2}$. Using the transformation 
properties (\ref{hp}), one finds that the constant is zero.

Our second example is the curve
\beq\label{C5}
R^2(x^2y^2+1)-(x^2+y^2)+Vxy=0.
\eeq
Its uniformization reads
\beq\label{C6}
x=\frac{\theta_4(u)}{\theta_1(u)}, \qquad
y=\frac{\theta_4(u+\eta )}{\theta_1(u+\eta )},
\eeq
and
\beq\label{C7}
R=\frac{\theta_1(\eta )}{\theta_4(\eta )}, \qquad
V=2\frac{\theta_4^2(0)\theta_2(\eta )\theta_3(\eta )}{\theta_2(0)
\theta_3(0)\theta_4^2(\eta )}.
\eeq
To see this, we should prove the identity
\beq\label{C8}
\begin{array}{c}
\displaystyle{
\frac{\theta_1^2(\eta )}{\theta_4^2(\eta )}\left (
\frac{\theta_4^2(u)\theta_4^2(u+\eta )}{\theta_1^2(u)\theta_1^2(u+\eta )}
+1\right ) -\left (\frac{\theta_4^2(u)}{\theta_1^2(u)}+
\frac{\theta_4^2(u+\eta )}{\theta_1^2(u+\eta )}\right )}
\\ \\
\displaystyle{
+2\, \frac{\theta_4^2(0)\theta_2(\eta )\theta_3(\eta )\theta_4(u)
\theta_4 (u+\eta )}{\theta_2(0)\theta_3(0)\theta_4^2(\eta )
\theta_1(u)\theta_1(u+\eta )}=0}.
\end{array}
\eeq
The left hand side is an elliptic function of $u$ with possible poles
at $u=0$ and $u=-\eta$ up to the second order. It is easy to see that 
the highest singularities (poles of the second order) cancel.
Therefore, the left hand side is an elliptic function of $u$ with
possible simple poles at $u=0$ and $u=-\eta$. Therefore, it is enough
to establish the equality at 3 distinct points. It is easy to see that
the left hand side equals 0 at $u=\frac{\tau}{2}$ 
and $u=-\eta +\frac{\tau}{2}$. Let the third point 
be $u=\frac{\tau +1}{2}$. At this point the left hand side is
$$
\mbox{l.h.s.}=\frac{\theta_1^2(\eta )}{\theta_4^2(\eta )}
\left (\frac{\theta_2^2(0)\theta_2^2(\eta )}{\theta_3^2(0)\theta_3^2(\eta )}
+1\right ) -\left (\frac{\theta_2^2(0)}{\theta_3^2(0)}+
\frac{\theta_2^2(\eta )}{\theta_3^2(\eta )}\right )+2\,
\frac{\theta_4^2(0)\theta_2^2(\eta )}{\theta_3^2(0)\theta_4^2(\eta )}.
$$
It is an even elliptic function of $\eta -\frac{\tau}{2}$ and 
$\eta -\frac{\tau +1}{2}$ with possible second order poles at $\eta =
\frac{\tau}{2}$ and $\eta =\frac{\tau +1}{2}$. The expansion around 
these points shows that the singular terms cancel. Therefore, the 
l.h.s. does not depend on $\eta$. Substituting $\eta =0$, we see that
this expression is equal to zero. This proves the identity (\ref{C8}).

\section*{Acknowledgments}

\addcontentsline{toc}{section}{Acknowledgments}

We thank A.Sleptsov who suggested the problem.

\end{document}